\documentclass[fleqn,usenatbib]{mnras}

\usepackage{newtxtext,newtxmath}

\usepackage[T1]{fontenc}

\DeclareRobustCommand{\VAN}[3]{#2}
\let\VANthebibliography\thebibliography
\def\thebibliography{\DeclareRobustCommand{\VAN}[3]{##3}\VANthebibliography}

\newcommand{\ADcomment}[1]

\newcommand{\HLcomment}[1]

\usepackage{graphicx}	
\usepackage{multirow}
\usepackage{booktabs}
\usepackage{float}
\usepackage{subfigure}
\usepackage{amsmath}	
\usepackage{longtable}
\usepackage{CJKutf8, color}

\title{EOS-dependent millihertz quasi-periodic oscillation in low-mass X-ray binary}
\author[Helei Liu et al.]{
Helei Liu,$^{1}$\thanks{E-mail: heleiliu@xju.edu.cn}
Yong Gao,$^{2,3}$
Zhaosheng Li,$^{4}$
Akira Dohi,$^{5,6}$
Weiyang Wang,$^{2,3}$
Guoliang L\"{u}$^{7,1}$
and Renxin Xu $^{2,3}$\thanks{E-mail: r.x.xu@pku.edu.cn}
\\
$^{1}$School of Physical Science and Technology, Xinjiang University, Urumqi 830046, China  \\
$^{2}$Department of Astronomy, Peking University, Beijing 100871, China\\
$^{3}$Kavli Institute for Astronomy and Astrophysics, Peking University, Beijing 100871, China\\
$^{4}$Key Laboratory of Stars and Interstellar Medium, Xiangtan University, Xiangtan 411105, Hunan, China\\
$^{5}$Astrophysical Big Bang Laboratory (ABBL), RIKEN Cluster for Pioneering Research, 2-1 Hirosawa, Wako, Saitama 351-0198, Japan\\
$^{6}$RIKEN Interdisciplinary Theoretical and Mathematical Sciences Program (iTHEMS), 2-1 Hirosawa, Wako, Saitama 351-0198, Japan\\
$^{7}$Xinjiang Astronomical Observatory, Chinese Academy of Science, 150 Science 1-Street, Urumuqi 830011, China
}

\date{Accepted XXX. Received YYY; in original form ZZZ}

\pubyear{2015}

\begin{document}
\label{firstpage}
\pagerange{\pageref{firstpage}--\pageref{lastpage}}
\maketitle

\begin{abstract}
We studied the frequency and critical mass accretion rate of millihertz quasi-periodic oscillations (mHz QPOs) using a one-zone X-ray burst model. The surface gravity is specified by two kinds of equation of states: neutron star (NS) and strange star (SS). The base flux, $Q_{b}$, is set in the range of 0-2 MeV nucleon$^{-1}$. It is found that the frequency of mHz QPO is positively correlated to the surface gravity but negatively to the base heating. The helium mass fraction has a significant influence on the oscillation frequency and luminosity. The observed 7-9 mHz QPOs can be either explained by a heavy NS/light SS with a small base flux or a heavy SS with a large base flux. As base flux increases, the critical mass accretion rate for marginally stable burning is found to be lower. Meanwhile, the impact of metallicity on the properties of mHz QPOs was investigated using one-zone model. It shows that both the frequency and critical mass accretion rate decrease as metallicity increases. An accreted NS/SS with a higher base flux and metallicity, combined with a lower surface gravity and helium mass fraction, could be responsible for the observed critical mass accretion rate ($\dot{m}\simeq 0.3\dot{m}_{\rm Edd}$). The accreted fuel would be in stable burning if base flux is over than $\sim$2 MeV nucleon$^{-1}$. This finding suggests that the accreting NSs/SSs in low-mass X-ray binaries showing no type I X-ray bursts possibly have a strong base heating.
\end{abstract}




\section{Introduction}

It has a long history to understand dense matter with both sub-nuclear and supra-nuclear densities~\citep{Fowler1926}, and a state of neutron star (NS) matter with extremely low charge-mass-ratio was proposed by \cite{Landau1932} before the discovery of neutron~\citep{Yakovlev2013}.
However, after the establishment of the standard model of particle physics, another way to make a neutral state of matter has also been focused~\citep[e.g.,][for a brief review]{Xu2023}, resulting in so-called strange matter with symmetry restoration of 3-flavored quarks (i.e., $u$, $d$ and $s$).
It is then a general thought that strangeness
would play an essential role in understanding the equation of state (EOS) of dense matter in pulsar-like compact objects.
As for the basic units of supra-nuclear matter with strangeness, quarks could be itinerant there if the running coupling of the fundamental strong interaction is asymptotically free~\citep{1986ApJ...310..261A,Haensel1986,1984PhRvD..30..272W}, but it is as well speculated that quarks could be clustered in nucleon-like units (so-called strangeons) if rich non-perturbative effects dominate at a few nuclear densities~\citep{2003ApJ...596L..59X}.
For the sake of simplicity, we call them strange stars (SSs) in this paper, to be independent of which form the quarks take (i.e., either quarks or strangeons).
Nevertheless,  thermonuclear X-ray bursts on compact star's surface are test-bed for us to identify the EOS, and this is our focus of the present work.

Type I X-ray bursts are powered by unstable thermonuclear burning of accumulated hydrogen and helium on the surface of NS in a low-mass X-ray binary~\citep[LMXB; for reviews, see][]{1993SSRv...62..223L,1998ASIC..515..419B,2006csxs.book..113S,2021ASSL..461..209G}. The mass accretion rate $\dot{M}$ is thought to determine the different burning regimes (e.g.\cite{1981ApJ...247..267F,1998ASIC..515..419B} ). Either theoretical arguments or observational evidence indicate that thermonuclear burning of the accreted material proceeds stably at very high $\dot{M}$, close to or above the Eddington mass accretion rate $\dot{M}_{\rm Edd}$~\citep{1993SSRv...62..223L,1998ASIC..515..419B}. \cite{1983ApJ...264..282P} 
predicted an oscillatory burning regime near the transition between unstable and stable burning, which is also known as marginally stable burning.
At present we know of 118 bursting sources\footnote{https://personal.sron.nl/$\sim$ jeanz/bursterlist.html}. A much smaller number of objects, ten at present count, have shown millihertz quasi-periodic oscillations (mHz QPOs). 

\cite{2001A&A...372..138R} reported the first detection of mHz QPOs in three NS low-mass X-ray binaries: 4U 1608-52, 4U 1636-53 and Aql X-1. These mHz QPOs occur only within a narrow range of luminosity, $L_{X}\approx0.5-1.5\times10^{37}\,\rm erg\,s^{-1}$, in the frequency range 7-9 mHz. They attributed this new phenomenon to a special mode of nuclear burning on the NS surface. 
\cite{2007ApJ...665.1311H} found that marginally stable nuclear burning on the NS surface could lead to the observed mHz QPOs. 

There are discrepancies between the observations and theory, for instance, the disagreement in the accretion rate for the oscillation to exist. Theoretical models of hydrogen and helium burning suggest this critical mass accretion rate to be located around the Eddington limit, while observations place the marginally stable burning close to $0.1\,\dot{M}_{\rm Edd}$. \cite{2007ApJ...665.1311H} proposed that the local accretion rate at the burning depth, where the mHz QPOs happen, can be higher than the entire NS surface. \cite{2007ApJ...663.1252P} and \cite{2009A&A...502..871K} showed a possible solution to solve this discrepancy by including mixing processes, e.g. rotation and rotationally induced magnetic fields. Furthermore, \cite{2014ApJ...787..101K} showed that the range of accretion rates with marginally stable burning depends strongly on both the composition of the burning layer and reaction rates.

Later, these mHz QPOs were detected in seven other sources, IGR J17480-2446~\citep{2010ATel.2958....1L}, 4U 1323-619~\citep{2011ATel.3258....1S}, IGR J00291+5934~\citep{2017MNRAS.466.3450F}, GS1826-238~\citep{2018ApJ...865...63S}, EXO 0748-676~\citep{2019MNRAS.486L..74M}, 1RXS J180408.9-342058~\citep{2021MNRAS.500...34T} and 4U 1730-22~\citep{2023MNRAS.521.5616M}. The mHz QPOs have frequencies below $\sim 14$ mHz. These mHz QPOs are evidently related to the thermal nuclear burning on the NS surface, as type I X-ray bursts are also present in the observations.

Some mHz QPOs appear with different characteristics. The persistent luminosity of IGR J17480-2446 when the mHz QPOs were observed was high, $L_{2-50~\rm {kev}}\sim 10^{38}~ \rm erg~s^{-1}$, the frequency was always below 4.5 mHz ~\citep{2010ATel.2958....1L}. \cite{2008ApJ...673L..35A} observed frequency drifts in mHz QPOs from 4U 1636-53. \cite{2015MNRAS.454..541L} measured the frequency drift time of the mHz QPOs in 4U 1636-536 from XMM-Newton and Rossi X-ray Timing Explorer and speculated that the drift timescale was set by cooling of the deeper layers. \cite{2021ApJ...920...35L} discovered a 2.7-11.3 mHz QPOs after a superburst in Aql X-1.

Understanding the observations of mHz QPOs allows us to probe the envelope composition, nuclear reaction rate, and the NS EOSs~\citep{2007ApJ...665.1311H,2014ApJ...787..101K,2021ApJ...923...64D,2023ApJ...950..110Z}. No matter the one-zone model or the multi-zone model such as KEPLER for theoretical calculations of thermonuclear burning, a base flux $Q_{\rm b}$ is included as an artificial parameter, which models the heat flowing from the underlying NS crust into the envelope. Most simulations use a low value of $Q_{\rm b}=0.1-0.15\,\rm MeV ~u^{-1}$~\citep{2007ApJ...665.1311H,2018ApJ...860..147M}. However, more recent work indicates that the base flux may be larger, up to $Q_{\rm b}=2\,\rm MeV\, u^{-1}$ ~\citep{2003A&A...404L..33H,2007ApJ...662.1188G}. Additionally, a yet unknown shallow heat source may increase the base flux~\citep{2006ApJ...646..429C,2015ApJ...809L..31D,2018PhRvC..98b5801F,2021PhRvD.104l3004L}. Furthermore, a SS can also have a crust of normal baryonic matter~\citep{1986ApJ...310..261A}. ~\cite{2005ApJ...635L.157P} studied the superburst from SS by considering the energy release from the dripped neutrons fall into quark matter. How about the mHz QPOs from the marginally stable burning on accreting SSs? The gravity and base flux may cause the differences compared with the accreting NSs.

In this work, we investigate the dependence of the oscillation frequency and critical mass accretion rate on the gravity and base flux using one-zone X-ray burst model. The gravity will be obtained from the EOSs of typical NS, SS. In Section ~\ref{sec:model}, we describe the basic equations of the one-zone model, the adopted EOSs and inputting parameters. In Section ~\ref{sec:QPOs}, we show the results of the marginally stable burning from NS and SS. The effect of gravity, base flux and helium mass fraction on the frequency and critical mass accretion rate are discussed. Besides, the impact of metallicity on the properties of mHz QPOs with use of one-zone model are investigated. In section ~\ref{sec:GR}, the general relativity corrections are discussed. Section ~\ref{sec:con} is devoted to conclusion.

\section{One-zone model }\label{sec:model}
To study the marginally stable burning, we use a publicly available code of one-zone model by A. Cumming\footnote{https://github.com/andrewcumming/onezone} (see also ~\cite{1983ApJ...264..282P,2007ApJ...665.1311H}). We describe the basic equations and micro-physics considered in the following calculations.
\subsection{The basic equations}
For the one-zone model calculations, the basic equations are as follows~\citep{1998ASIC..515..419B,2007ApJ...665.1311H}:
\begin{eqnarray}
   P &=& g_{\rm s}y, \label{eq:hyd}\\
   c_p\frac{dT}{dt} &=&  \varepsilon-\frac{F}{y}, \label{eq:bal}\\
   \frac{dy}{dt} &=& \dot{m}-\frac{\varepsilon}{E_{*}}y. \label{eq:tra}
\end{eqnarray}\\
where $g_{\rm s}$ is the surface gravity, $c_{p}$ is the specific heat capacity at the constant pressure, and $y$ is the column density, in units of mass per unit area, $\dot{m}$ is the local accretion rate. $\varepsilon$ is the heating from the burning layer, the code considers three kinds of energy generation: the energy release from the triple-$\alpha$ reaction, hot-CNO hydrogen burning and base heating from the bottom of the accumulating fuel layer. $E_{*}$ is the energy per gram released in the burning layer, $E_{*}=5.3\,\rm MeV/u$ is adopted in this work, where the changes of $E_{*}$ with hydrogen mass fraction are not considered here \,\citep{2007ApJ...665.1311H, 1999ApJ...524.1014S}. Equation (\ref{eq:hyd}) is derived from hydrostatic balance, which describes the pressure $P$ at the base of the accretion layer. Equation (\ref{eq:bal}) describes the heat balance, including the energy generation rate $\varepsilon$, and radiative loss rate, $-\nabla \cdot \textbf{F}/\rho=dF/dy\approx F/y$. The flux $F=acT^4/3\kappa y$~\citep{1998ASIC..515..419B}, where the opacity $\kappa$ is calculated as described by ~\cite{1999ApJ...524.1014S}.
Equation (\ref{eq:tra}) tracks the burning depth. The 
pressure $P$ includes the non-relativistic degenerate electron pressure, ion pressure and radiation pressure, the details of the equation of state in the accreting layer can be referred to Equation (24) from ~\cite{1983ApJ...264..282P}.
In the following, we will study the marginally stable burning by using the public code \texttt{onezone}.

\subsection{The EOS and heating}
To date, there is no real evidence against or in favor of the SS hypothesis ~\citep{2003NuPhA.715..835P}, since the properties such as radius and moment of inertia are similar for NS and SS in the observed mass range $1\,M_{\odot}<M<2\,M_{\odot}$. The X-ray burst phenomenon is modeled on the surface of accreting NSs in most of the works. However, there are many works support the existence of the accreting SS~\citep{1991NuPhS..24..139H,2005ApJ...635L.157P,2013PASP..125...25Z,2018ApJ...858...88Z}. If we consider the X-ray burst from the surface of accreting SS, there will be $\sim 10-100\,\rm MeV$ of energy that comes from dripped neutrons fall into the quark matter~\citep{1984PhRvD..30.2379F,2005ApJ...635L.157P}. Much of this energy is carried off by neutrino, but a considerable part of this heat will be left and alter the temperature of the burning layer. ~\cite{2005ApJ...635L.157P} considered this energy and calculated the ignition condition for the superburst from SSs. The consequences of this alteration for thermal nuclear burst model have not been explored. On the other hand, the cold quark matter could be strange quark clusters because of the strong interaction between quarks~\citep{2003ApJ...596L..59X}. A strange quark cluster is named 'strangeon', being coined to `strange nucleon' for simplicity~\citep{2017JPhCS.861a2027X}. Therefore, the pulsar like compact stars could be strangeon stars. How about X-ray burst on the accreting strangeon star?

In this work, we simplified the X-ray burst simulations from accreting NS, SS with use of one-zone model. The different properties of them are reified by the surface gravity which is obtained from the mass-radius ($M-R$) relationship and base flux. Thus, EOS and base flux are important factors to determine $g_{\rm s}$ and energy generation rate which distinguish the X-ray burst scenario. For NS, we adopt the Togashi EOS which was constructed by using realistic two-body potential and phenomenological three-body potential~\citep{2017NuPhA.961...78T}. For SS star, we consider strange quark star and strangeon star, for the former one, the MIT bag model of quark matter with bag constant $B=70 ~\rm MeV~fm^{-3}$ is used~\citep{1986ApJ...310..261A}. For the latter one, we adopted LX3630 EOS which is based on the Lennard-Jones model with the surface baryon number density $n_{\rm s}=0.36~\rm fm^{-3}$ and the potential depth $\epsilon=30~\rm MeV$~\citep{2022MNRAS.509.2758G,2009MNRAS.398L..31L}.

\begin{figure*}
    \centering
    \begin{minipage}{0.49\linewidth}
    \centering
         \includegraphics[width=0.9\linewidth]{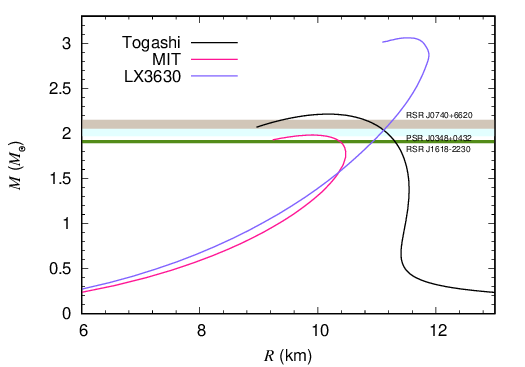}
         \end{minipage}
         \begin{minipage}{0.50\linewidth}
         \includegraphics[width=\linewidth]{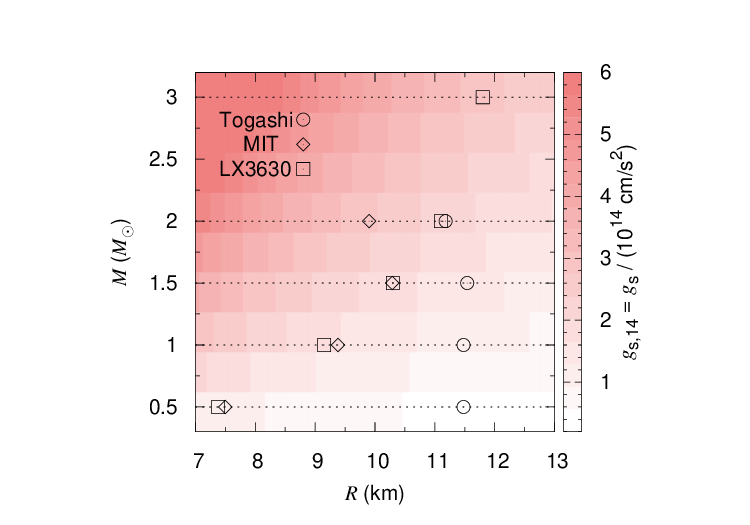}   
         \end{minipage}
    
    \caption{Left: $M-R$ relations of NS (black), strange star (pink) and strangeon star (purple) adopted in this work, and the $1\sigma$ region of the mass measurements of PSR J1618$-$2230~\citep{2010Natur.467.1081D}, PSR J0348+0432~\citep{2013Sci...340..448A}, PSR J0740+6620~\citep{2021ApJ...915L..12F} are highlighted. Right: The surface gravity acceleration $g_{\rm s}$ on the $M-R$ plane. The symbols indicate the values of mass and radius of NSs/SSs in our model calculations.}
    \label{fig:mr}
\end{figure*}

The $M-R$ relations of the three adopted EOSs are shown in the left panel of Fig.\ref{fig:mr}. In the right panel of Fig.\ref{fig:mr}, we show the values of the local gravity $g_{\rm s}$ obtained using the Newtonian formula $g_{\rm s}={GM}/{R^2}$ for several $M$ and $R$. However, note that general relativity corrections should be applied when relating a value of $g_{\rm s}$ to $M$ and $R$ from a particular EOS and when using the results for mHz QPOs (see discussion in section 4.4 of \cite{2004ApJS..151...75W} and Section~\ref{sec:GR}). Nonetheless, \cite{2007ApJ...665.1311H} noted that \texttt{onezone} code overpredicts the oscillation frequency, thus somewhat compensating the corrections (see footnote 7 of~\cite{2007ApJ...665.1311H}).
In table~\ref{tab:g14}, we show the masses and corresponding (Newtonian) surface local gravity for each EOSs, where $g_{14}$ is the gravitational acceleration in unites of $10^{14}~\rm cm/s^2$.

For the nuclear burning, the code considers hydrogen and helium burning ~\citep{2007ApJ...665.1311H}. The nuclear energy generation rate from the hot CNO hydrogen burning is as follows~\citep{1981ApJS...45..389W}:
\begin{eqnarray} \label{eq:cno}
    \varepsilon_{\rm hCNO}=5.86\times10^{15} Z~\rm {erg~g^{-1}~s^{-1}}
\end{eqnarray}
where $Z$ represents the sum of the mass fraction of the metallicity inside the accreted layer. For the helium burning, we use the $3\alpha$ reaction rate as given by ~\cite{1987ApJ...317..368F}. The code includes a factor of $E_{\rm nuc}/E_{3\alpha}$ to enhance the $3\alpha$ reaction rate ~\citep{2007ApJ...665.1311H}, where $E_{\rm nuc}=1.6+4.9\,X_{0}~\rm MeV/u$, $X_{0}$ is the mass fraction of hydrogen in the accreted layer. $E_{3\alpha}=0.606\,\rm MeV/u$ is the energy release from the $3\alpha$ reaction. Besides, the code allows for the inclusion of a base heating at the bottom of the accreted layer. The values of $Q_{\rm b}$ is set in the range of $0-2\,\rm MeV/u$, the corresponding base luminosity $L_{\rm b}=Q_{\rm b}\dot{M}$. When $Q_{\rm b}>2\,\rm MeV/u$, the accreted fuel will be in stable burning which we will discuss in the next section. The total energy generation rate $\varepsilon$ contains the nuclear burning including hydrogen and helium and the base heating. We use the public \texttt{onezone} code to integrate equations (\ref{eq:bal}) and (\ref{eq:tra}) with respect to time to determine the evolution of the burning layer. The simulations are started with an arbitrary conditions at $T=2\times 10^8\, \rm K$ and $y=2\times 10^8\,\rm g\,cm^{-2}$ following ~\cite{2007ApJ...665.1311H}.

\begin{table}
    \centering
    \caption{The values of local gravity for several values of $M$ with different EoSs.}
    \begin{tabular}{c|ccc}
    \hline\hline
    Mass ($M_{\odot}$) & $g_{14}(\rm Togashi)$ & $g_{14}(\rm MIT)$ & $g_{14}(\rm LX3630)$\\
    \hline
    0.5 & 0.50 & 1.15 & 1.22\\
    \hline
    1.0 & 1.01 & 1.51 & 1.59\\
    \hline
    1.4 & 1.39 & 1.80 & 1.82\\
    \hline
    1.5 & 1.50 & 1.88 & 1.88\\
    \hline
    2.0 & 2.12 & 2.69 & 2.16\\
    \hline
    3.0 & - & - & 2.86\\
     \hline\hline 
    \end{tabular}
      \label{tab:g14}
\end{table}

\section{Marginally stable burning from accreting NS and SS }\label{sec:QPOs}
The mHz QPO phenomenon is thought to occur at the boundary of unstable and stable nuclear burning. This oscillatory model of burning also called "marginally stable burning". ~\cite{2007ApJ...665.1311H} studied the marginally stable nuclear burning on an accreting NS with both one-zone model and multi-zone hydrodynamic KEPLER code.  They derived the period of the oscillations close to 2 minutes, which is in good agreement with the range of oscillation frequencies reported for the mHz QPOs. Here, we consider the marginally stable burning from accreting NS and SS with one-zone model as described in section ~\ref{sec:model}. For convenient, we define the local Eddington accretion rate $\dot{m}_{\rm Edd}\equiv8.8\times 10^4~\rm g~cm^{-2}~s^{-1}$ throughout this paper.
\subsection{Light curves for the one-zone model from accreting NS and SS}
\label{sec:maths} 

\begin{figure}
    \centering
    \includegraphics[width=\columnwidth]{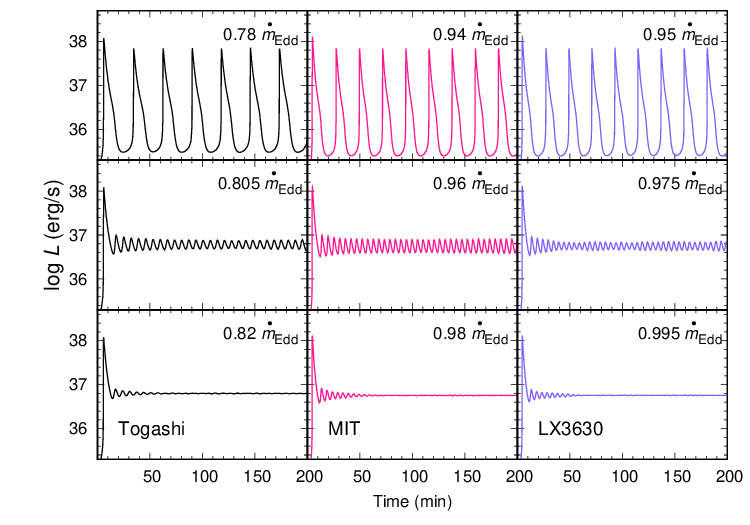}
    \caption{Light curves for the one-zone model at three different accretion rates, where $M=1.4\,M_{\odot}$ and $Q_{\rm b}=0.15\,\rm MeV/u$, $X_0=0.7$. For Togashi EoS, at $\dot{m}=0.78\,\dot{m}_{\rm Edd}$, the recurrence time of bursts is 28 minutes.  At $\dot{m}=0.805\,\dot{m}_{\rm Edd}$, oscillations are seen with a period of 6.8 minutes. At $\dot{m}=0.82\,\dot{m}_{\rm Edd}$, after a few transient oscillations, the burning evolves to a steady state. For MIT EoS, at $\dot{m}=0.94\,\dot{m}_{\rm Edd}$, the recurrence time of bursts is 22 minutes.  At $\dot{m}=0.96\,\dot{m}_{\rm Edd}$, oscillations are seen with a period of 5.2 minutes. The burning evolves to a steady state at $\dot{m}=0.98\,\dot{m}_{\rm Edd}$. For LX3630 EoS, at $\dot{m}=0.95\,\dot{m}_{\rm Edd}$, the recurrence time of bursts is 22 minutes.  At $\dot{m}=0.975\,\dot{m}_{\rm Edd}$, oscillations are seen with a period of 5.2 minutes. The burning evolves to a steady state at $\dot{m}=0.995\,\dot{m}_{\rm Edd}$.  }
    \label{fig:lc}
\end{figure}

Figure\,\ref{fig:lc} shows light curves from the one-zone model at different accretion rates. We set the local gravity to be the Newtonian gravity for a $1.4\,M_{\odot}$ NS/SSs star from Togashi, MIT and LX3630 EoSs, the corresponding $g_{14}$ equals 1.39, 1.80 and 1.82, where $g_{14}=g_{\rm s}/(10^{14} \rm \,cm^{2}\,s^{-1})$. We show light curves at $\dot{m}=0.78\,\dot{m}_{\rm Edd}, 0.805\,\dot{m}_{\rm Edd}, 0.82\,\dot{m}_{\rm Edd}$ for Togashi EoS, $\dot{m}=0.94\,\dot{m}_{\rm Edd}, 0.96\,\dot{m}_{\rm Edd}, 0.98\,\dot{m}_{\rm Edd}$ for MIT EoS, $\dot{m}=0.95\,\dot{m}_{\rm Edd}, 0.975\,\dot{m}_{\rm Edd}, 0.995\,\dot{m}_{\rm Edd}$ for LX3630 EoS. Close to the stability boundary, the recurrence time is 28 minutes for Togashi EoS, 22 minutes for MIT and LX3630 EoSs. At accretion rates above the boundary, the burning evolves to a steady state. At $\dot{m}=0.805\,\dot{m}_{\rm Edd}, 0.96\,\dot{m}_{\rm Edd}, 0.975\,\dot{m}_{\rm Edd}$, we see oscillations with an oscillation period of 6.8 minutes, 5.2 minutes and 5.2 minutes for Togashi EoS, MIT EoS and LX3630 EoS, respectively. The more compact the star, the greater the transition mass accretion rate.

\begin{figure*}
    \centering
    \begin{minipage}{0.49\linewidth}
    \centering
         \includegraphics[width=\linewidth]{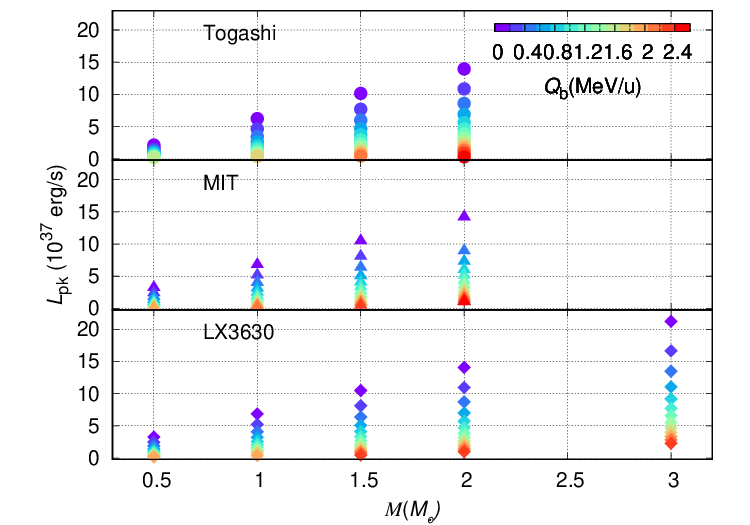}
         \end{minipage}
         \begin{minipage}{0.49\linewidth}
         \includegraphics[width=\linewidth]{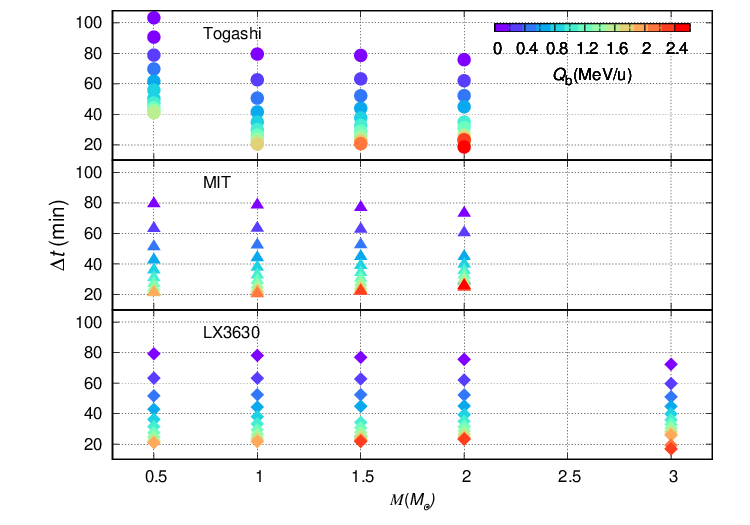}   
         \end{minipage}
    
    \caption{The peak luminosity and recurrence time of the bursts for a series NS/strange(strangeon) star. The different colours mark the different values of the base flux. The mass accretion rate is set as $\dot{m}=0.3\,\dot{m}_{\rm Edd}$, $X_0=0.7$. }
    \label{fig:changes}
\end{figure*}
Figure\,\ref{fig:changes} shows the peak luminosity and recurrence time change with masses and base flux $Q_{\rm b}$. The bursts are simulated at a fixed mass accretion rate $\dot{m}=0.3\,\dot{m}_{\rm Edd}$ and hydrogen fraction $X_0=0.7$. It shows a positive correlation for every EOS for the peak luminosity at a fixed $Q_{\rm b}$, but an anti-correlation for every EOS for the recurrence time at a fixed $Q_{\rm b}$. At a fixed mass for every EOS, both the peak luminosity and recurrence time are anti-correlation with $Q_{\rm b}$. The accreted material will be in steady burning if the base flux $Q_{b}$ is high, e.g., for $0.5\,M_{\odot}$ NS, the critical $Q_{b}\simeq1\,\rm MeV/u$, for $2.0\,M_{\odot}$ NS, the critical $Q_{b}\simeq2.1\,\rm MeV/u$, for $3.0\,M_{\odot}$ strangeon star\footnote{The $3\,M_{\odot}$ SS is adopted to show the effect of strong gravity. However, the mass measurement of the three massive pulsars in the left panel of Figure\,\ref{fig:mr}  show that the maximum mass of pulsar is about $\sim 2\,M_{\odot}$. In order to compare with the massive mass measurement of pulsar, we thought that the accreted fuel would be in stable burning if $Q_{\rm b}\gtrsim 2\,\rm MeV/u$ for simplicity.}, the critical $Q_{b}\simeq2.5\,\rm MeV/u$. 

\subsection{Oscillation frequency for the one-zone model from accreting NS and SS}
\begin{figure}
    \centering
    \includegraphics[width=\columnwidth]{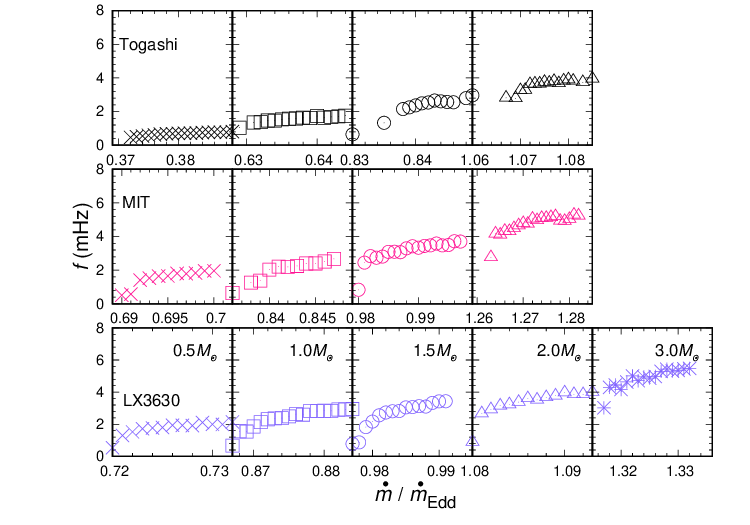}
    \caption{Oscillation frequency as a function of mass accretion rate for different choices of NS/strange(strangeon) star masses. $X_0=0.7$, $Q_{\rm b}=0.15~\rm MeV/u$.}
    \label{fig:tmdot}
\end{figure}

In the following, we quantify the properties of the oscillations in the light curves of all simulations with marginally stable burning by determining the oscillation period/frequency, and the maxima and minima luminosity. 

Figure\,\ref{fig:tmdot} shows the dependence of the (uncorrected) oscillation frequency on the accretion rate for different choices of NS/SS masses. At each accretion rate, we integrate the one-zone model for $10^6~\rm s$, and plot the average oscillation frequency after discarding the first 100 minutes of data following ~\cite{2007ApJ...665.1311H}. We can see that the overall range of accretion rates for which oscillations are seen is very narrow, $\Delta \dot{m}\approx 0.01\dot{m}_{\rm Edd}$. For $0.5\,M_{\odot}$ NS, it shows a very low frequency mHz QPOs, $f<1$ mHz, although we have not observed such a low frequency mHz QPOs so far. For a $0.5\,M_{\odot}$ strange(strangeon) star, the oscillation frequency is $\sim 2~\rm mHz~QPOs$. For a higher mass NS/SS, as the surface gravity increases, the oscillation frequency increases. For $1.5\,M_{\odot}$ NS and SS, the oscillation frequencies are $\sim 3$ mHz and $\sim 4$ mHz, respectively. For $2.0\,M_{\odot}$ NS and SS, the oscillation frequencies are $\sim 4$ mHz and $\sim 5$ mHz, respectively. For the same mass NS and SS, the oscillation frequency is higher for SS star. Our results are consistent with Figure 4 of ~\cite{2007ApJ...665.1311H}.

\begin{figure}
    \centering
    \includegraphics[width=\columnwidth]{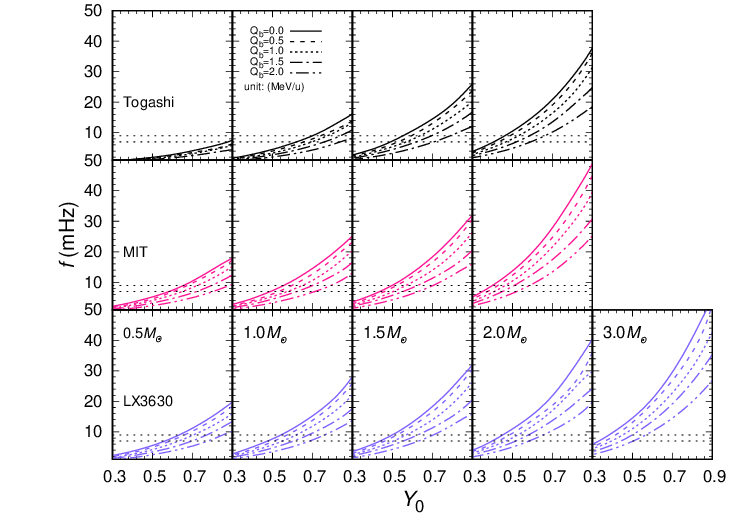}
    \caption{Oscillation frequency as a function of the helium mass fraction, $Y_0$, for different choices of NS/SS masses and base flux, where the metallicity is set as $Z=0.01$. The horizontal dotted lines in each panel show the boundaries of the first observed 7-9 mHz QPOs.}
    \label{fig:Of}
\end{figure}

One-zone model has been used to explore how the properties of the oscillations change with parameters such as surface gravity and accretion composition~\citep{2007ApJ...665.1311H}. As the surface gravity is degeneracy between the combinations of mass and radius, the previous works do not distinguish the gravity from NS and SS. Here we adopt one-zone model, but the surface gravity is inferred from three representative EoSs. The effect of base heating $Q_{\rm b}$ on oscillation properties also will be investigated.
 Compared with multi-zone model coupled with the nuclear reaction network, one-zone model saves a lot of computation time. We have performed simulations for variations of helium mass fraction and base flux for a series of NS/SS masses, where the metallicity is set as $Z=0.01$.

Figure\,\ref{fig:Of} shows the dependence of the oscillation frequency on the helium mass fraction of the accreted material and base flux for several NS/SS masses. When increasing $Y_0$, the oscillation frequency increases. Values of $f$ range from $0.35\,\rm mHz$ to $\sim39\,\rm mHz$ for NS, $0.84\,\rm mHz$ to $\sim54\,\rm mHz$ for SS. The oscillation frequency decreases as $Q_{\rm b}$ increases,  increases as the mass of NS/SS increases.  For the same NS/SS mass, the oscillation frequency of SS is higher than NS if the base flux and hydrogen mass fraction are also the same. For NS, the 7-9\,mHz QPOs can be fitted with a $1.5\,M_{\odot}$ with helium mass fraction $Y_0$ in the range of 0.6-0.7, while for SS, the 7-9\,mHz QPOs can be satisfied with a $1.5\,M_\odot$ with a helium mass fraction $Y_0$ in the range of 0.5-0.6. The details of the output quantities can be found in Tables 1-3 in Appendix.

\begin{figure}
    \centering
    \includegraphics[width=\columnwidth]{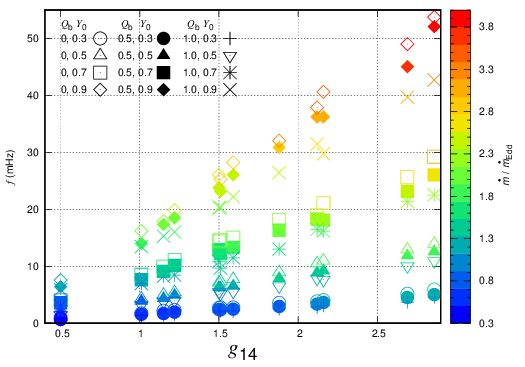}
    \caption{Oscillation frequency vs. surface gravity for different base heating $Q_{\rm b}$ and helium mass fraction $Y_0$ (marked as kinds of symbols). The colours indicate the critical mass accretion rate at which marginally stable burning occurs for each scenario.}
    \label{fig:fg}
\end{figure}

For above series of simulations, we study the properties of oscillation frequency as well as critical mass accretion rate with different NS/SS masses, accretion compositions and base flux. As we can get the surface gravity from the NS/SS's $M-R$ relationship ( the values of the local surface gravity we adopted can be seen from Table~\ref{tab:g14}), we show the distributions of the oscillation frequencies and surface gravity for different base heating and helium mass fractions scenarios in Figure~\ref{fig:fg}. The color represents the last value of the critical mass accretion rate in its narrow range at which the burning is marginally stable. For these similar models for $Q_{\rm b}$, $Y_0$ and $g_{14}$ variations, we find that surface gravity has a significant impact on oscillation frequency and critical mass accretion rate. At higher surface gravity, both the oscillations frequency and the critical mass accretion rate are larger. Moreover, the amplitudes of $f$ and $\dot{m}$ are higher at larger helium mass fraction and smaller base heating. Although our simulations are crude, the results suggest that one can infer the surface gravity if we obtain the frequency and mass accretion rate of the mHz QPOs.

\begin{figure*}
 
    \centering
    \includegraphics[width=1.8\columnwidth]{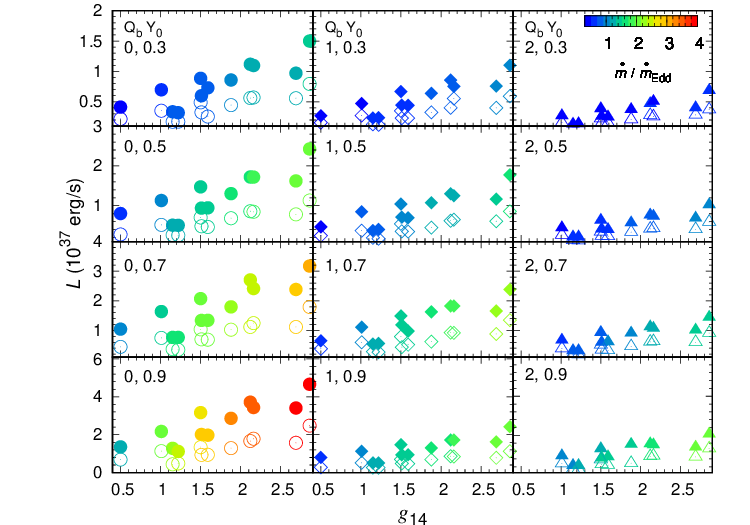}
    \caption{Oscillation luminosity vs. surface gravity for different base heating $Q_{\rm b}$ and helium mass fraction $Y_0$. The solid symbols in each panel indicate the maxima luminosity, the hollow symbols indicate the minima luminosity.}
    \label{fig:lg}
\end{figure*}

We specify the amplitude of the oscillations in the light curves of all simulations with marginally stable burning by determining the maxima and minima luminosity. Figure\,\ref{fig:lg} shows the distributions of the oscillation luminosity and surface gravity for different base heating and helium mass fraction models.
The colors of the symbols represent the critical mass accretion rate the same as Figure\,\ref{fig:fg}. The difference between the maxima luminosity and minima luminosity increases as surface gravity increases except some dips at several gravities ($g_{14}=1.22, 1.39,2.69$). Nevertheless, the trends in the properties are clearly: for larger $Q_{\rm b}$, the difference between the maxima luminosity and minima luminosity are smaller, while for larger $Y_0$, the difference between the maxima luminosity and minima luminosity are larger. Compared with the narrow luminosity observations, $\dot{m}/\dot{m}_{\rm Edd}\approx0.1-0.3$, we can exclude the models with low base flux and high helium mass fraction, e.g., the first panel in the fourth line of Figure~\ref{fig:lg}, for which the critical mass accretion rate is too large to fit the observations. Our models predict the critical mass accretion rate in the range $\sim0.27-4.3\,\dot{m}_{\rm Edd}$. Although observations place the critical mass accretion rate around $0.1\,\dot{m}_{\rm Edd}$ to $0.3\,\dot{m}_{\rm Edd}$ which is an order of magnitude lower than mass accretion rate implied by the model calculation, it has been suggested that rotational mixing of the accreted fuel to greater depths could alleviate this apparent discrepancy~\citep{2009A&A...502..871K}.

\subsection{The impact of metallicity on the properties of mHz QPOs}
From the above calculations, the oscillation properties have been studied by using one-zone model with variation in surface gravity, helium mass fraction and base heating. \cite{2014ApJ...787..101K} find that $Z$ and the nuclear reaction rates of the hot-CNO breakout reactions will change the properties of oscillation burning. However, \cite{2014ApJ...787..101K} adopt the multi-zone KEPLER code with a more complicated network, besides, Z is a linear function of hydrogen mass fraction $X_0$. So, how about the effect of Z for one-zone model? To answer this question, we study the effect of metallicity on the properties of mHz QPOs with use of one-zone model, where a typical NS ($1.5\,M_{\odot}$) with surface gravity $g_{14}=1.50$ and helium mass fraction $Y_0=0.3$ is adopted, $Z$ is set in the range of 0.01-0.1.

\begin{figure}
    \centering
    \includegraphics[width=\columnwidth]{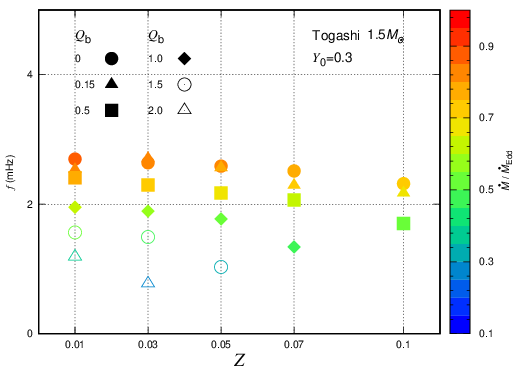}
    \caption{Oscillation frequency vs. metallicity for different base heating $Q_{\rm b}$ (in unit of MeV/u), where the helium mass fraction is set as $Y_0=0.3$, the surface gravity $g_{14}=1.50$ is inferred from a $1.5\,M_{\odot}$ NS (Togashi EoS). The colours indicate the critical mass accretion rate at which marginally stable burning occurs for each scenario.}
    \label{fig:z}
\end{figure}

Figure\,\ref{fig:z} exhibits the changes of oscillation frequency and critical mass accretion rate with the variation in metallicity. The symbols represent the different values of base heating, and the colors represent the values of critical mass accretion rate. Both the oscillation frequency and the critical mass accretion rate decrease as $Z$ increases. For $Q_{\rm b}=0$, the oscillation frequency decreases by $\thickapprox14\%$, the critical mass accretion rate decreases by $\thickapprox20\%$, for $Q_{\rm b}=0.5$, the oscillation frequency decreases by $\thickapprox29\%$, the critical mass accretion rate decreases by $\thickapprox32\%$. For $Q_{\rm b}=1.0\,\rm MeV/u$, it will be in stable burning when $Z>0.07$, and for $Q_{\rm b}=1.5\,\rm MeV/u$ and $Q_{\rm b}=2.0\,\rm MeV/u$, it will be in stable burning when $Z>0.05$ and $Z>0.03$, respectively. Thus, the high metallicity is helpful to explain the observation of low critical mass accretion rate.
For a fixed Z, the oscillation frequency and critical mass accretion rate decrease as $Q_{\rm b}$ increases. We show the details of the output quantities for the change of $Z$ in Table 4 in Appendix.

\subsection{Discussion}

SS differs from NS mainly in two ways. First, the $M-R$ relation for SS is very different from that of NS, especially, there is no minimum mass, and for $M\lesssim1\,M_{\odot}$, $M\varpropto R^3$. In contrast, NSs have radii that decrease with increasing mass. Second, accretion of matter onto the strange star surface will compress the matter at the base of the crust, inducing further electron captures on nuclei, followed by further neutron drip. Neutrons will be absorbed by the quark bag by the reactions $ n~\rightarrow ~u+2d$, $d+u~\rightarrow~u+s$, with heat release $\sim10-100\,$ MeV per one absorbed neutron~\citep{1984PhRvD..30.2379F,2005ApJ...635L.157P}.  However, accretion of matter onto the NS will compress the crust and induce the non-equilibrium processes including electron captures, neutron emission and absorption and pycnonuclear fusion, with heat release $\sim 1-2$\,MeV per one accreted nucleon~\citep{1990A&A...227..431H,2003A&A...404L..33H,2008A&A...480..459H}.
Since the properties of X-ray burst are determined by the mass, radius, accretion composition, base flux, nuclear reaction rate etc ~\citep{2018ApJ...860..147M,2020MNRAS.494.4576J,2021ApJ...923...64D,2022ApJ...937..124D}, the X-ray burst from accreting NS should be different from accreting SS due to the above mainly differences. Furthermore, ~\cite{2003ApJ...596L..59X} conjectured astrophysical cold quark matter could be in a sold state, because the strong interaction may render quarks grouped in clusters. Thus, strange quark star could be strangeon stars, X-ray burst is likely to occur on the surface of accreting strangeon star.

It's useful to study the differences of mHz QPOs from accreting NS/SS. As the surface gravity of SS is greater than NS for the same mass, the oscillation frequency is higher for SS. On the other hand, as the heat flux at the base of the crust from accreting SS is larger than NS, it's possible that the accreted fuel is in stable burning when falling onto the surface of the stellar, and it is more likely occurs on an accreting SS with a relatively thick crust (the base of the crust is at the neutron drip density). For the simple one-zone model, when $Q\gtrsim 2\,\rm MeV/u$, there will be no burst, the accreted fuel will be in stable burning. As far as we know, nearly half of all LMXBs show type I X-ray burst ~\citep{2004NuPhS.132..486I}, besides the LMXB which contains a black hole, SS star in LMXB is a good candidate that can not show thermonuclear burning. However, the accreting strange(strangeon) star with a relatively thin crust (the base of the crust has not reached neutron drip density) should be excluded, because there is no neutron drip, as a result, no energy release from the transition of neutron to quark. In this case, the X-ray burst from accreting SS is similar to accreting NS.

\section{Correcting for general relativity}\label{sec:GR}
To accurately model X-ray bursts, it is important to consider the effect of general relativity. \cite{2007ApJ...665.1311H} reached the conclusion about the fact that general relativity corrections were compensated by the simplified network of the one-zone model because they compared results between KEPLER and onezone codes that they were using. The above results are calculated from the local Newton frame, it's necessary for us to correct the results for general relativity~\citep{2004ApJS..151...75W}. In general relativity, the gravitational acceleration at the surface of the star is $g=GM_{\rm GR}(1+z)/R_{\rm GR}^2$, where the gravitational redshift is $z=(1-2GM_{\rm GR}/(R_{\rm GR}c^2))^{-1/2}-1$, $M_{\rm GR}$ and $R_{\rm GR}$ are the general relativity mass and radius, respectively. Assume that the surface gravity in Newtonian calculations is equal to that in the general relativity 
\begin{eqnarray} \label{eq:grcor}
    \frac{GM}{R^2}=\frac{GM_{\rm GR}(1+z)}{R_{GR}^2}
\end{eqnarray}
We can apply our results to any combination of $M_{\rm GR}$ and $R_{\rm GR}$. Here, we adopt the gravity in table \ref{tab:g14}, by fixing the original radii (assume $R=R_{\rm GR}$), we can obtain $M_{\rm GR}$ that correspond to those values of $g_{\rm s}$ by equation (\ref{eq:grcor}). The details of the values for correcting for general relativity are shown in table \ref{tab:GR}.

\begin{table}
    \centering
    \caption{The values for correcting for general relativity.}
    \begin{tabular}{c|cccc}
    \hline\hline
    Mass ($M_{\odot}$) & $M_{\rm GR}$ & R(km) & $g_{14}$ & 1+z\\
    \hline
    0.5 & 0.47 & 11.48 & 0.50 & 1.07\\
    \hline
    1.0 & 0.88 & 11.48 & 1.01 & 1.14\\
    \hline
    1.5 & 1.24 & 11.54 & 1.50 & 1.21\\
    \hline
    2.0 & 1.54 & 11.18 & 2.12 & 1.30\\
    \hline
    \hline
    0.5 & 0.45 & 7.61 & 1.15 & 1.10\\
    \hline
    1.0 & 0.85 & 9.38 & 1.51 & 1.17\\
    \hline
    1.5 & 1.21 & 10.29 & 1.88 & 1.24 \\
    \hline
    2.0 & 1.49 & 9.94 & 2.69 & 1.34 \\
    \hline
    \hline
    0.5 & 0.45 & 7.38 & 1.22 & 1.11\\
    \hline
    1.0 & 0.85 & 9.15 & 1.59 & 1.17\\
    \hline
    1.5 & 1.21 & 10.30 & 1.88 & 1.24\\
    \hline
    2.0 & 1.54 & 11.10 & 2.16 & 1.30\\
    \hline
    3.0 & 2.08 & 11.80 & 2.86 & 1.44\\
     \hline\hline 
    \end{tabular}
      \label{tab:GR}
\end{table}

In the observer's frame, the oscillation period should be increased by a factor $1+z$,
\begin{eqnarray}
    t_{\infty}=t(1+z)
\end{eqnarray}
As a result, the oscillation frequency will be divided by a factor $1+z$,
\begin{eqnarray}
    f_{\infty}=\frac{f}{1+z}
\end{eqnarray}
We show $f_{\infty}$ in tables 1-3 in the appendix.

\section{Conclusions} \label{sec:con}
To study the effects of EOS on mHz QPOs, we employ a set of simulations using a simple one-zone X-ray burst model. We obtained $g_{\rm s}$ from a specified NS/SSs, where Togashi, MIT and LX3630 EoSs are adopted in our simulations. The detailed studies show the differences of the properties of mHz QPOs due to $g_{\rm s}$, $Q_{\rm b}$, $Y_{0}$ and Z. The oscillation frequency from accreting  SS is higher than that from accreting NS for the same mass, since the surface gravity of SS is greater than NS for the same mass. For a very low mass NS/SS, e.g. $0.5\,M_{\odot}$, the oscillation frequency of accreting NS is less than 1 mHz when $Y_0=0.3$ , while for SS, the oscillation frequency is $\sim 2\,\rm mHz$. For a high mass NS and SS, e.g., $1.5\,M_{\odot}$, the oscillation frequencies are $\sim 3\,\rm mHz$ and $\sim 4\,\rm mHz$, respectively. For $2.0\,M_{\odot}$ NS and SS, the oscillation frequencies are $\sim 4\,\rm mHz$ and $\sim 5\,\rm mHz$, respectively.
  The Oscillation frequency and critical mass accretion rate decrease as base flux increases, the accreted fuel will be in stable burning when $Q_{\rm b}\gtrsim2\,\rm MeV/u$, which indicates that the accreting NS/SS in LMXBs showing no type I X-ray bursts may have a strong base flux. 

Meanwhile, we study the impact of accretion composition on mHz QPOs. The oscillation frequency increases as helium mass fraction increases. The oscillation frequency ranges from $\sim 0.93-10.5\,\rm mHz$ for NS, $\sim 0.85-13.9\,\rm mHz$ for SS if $Y_0$ is in the range $0.3-0.5$. The oscillation frequency ranges from $\sim 1.5-37.9\,\rm mHz$ for NS, $\sim 3.3-53.8\,\rm mHz$ for SS if $Y_0$ is in the range $0.5-0.9$. The critical mass accretion rate and oscillation amplitude increase as surface gravity and/or helium mass fraction increase, decrease as the base flux increases. Besides, the metallicity has non-negligible effect on the oscillation frequency and critical mass accretion rate. With increase of the metallicity, both the frequency and critical mass accretion rate decrease. Thus, the high metallicity and large base flux is helpful to reduce the discrepancy of critical mass accretion rate between models and observations. Since the above results are obtained from one-zone model with the local Newtonian gravity, we also correct the results for general relativity in section \ref{sec:GR}. 

So far, we have observed mHz QPOs from ten sources, for which the physical origin is considered as marginally stable burning. We have confirmed that the properties of these mHz QPOs are obviously affected by the $M-R$ relation and helium mass fraction. We also investigated the impacts of base heating and metallicity on the oscillation properties, and we found that the burst frequency and critical mass accretion rate will be decreaded as the base heating and metallicity increase. We may infer the surface gravity if we known the oscillation frequency and mass accretion rate of mHz QPOs from Figure~\ref{fig:fg}. It worth to mention that there are some high energy mHz QPOs which are different from the low photon energy ($\lesssim 5\,\rm keV$) mHz QPOs that discussed in this paper, e.g. 35-95\,mHz QPO is detected in the 27-120 keV energy band from the Be/X-ray binary 1A 0535+262 using Insight-HXMT data~\citep{2022MNRAS.517.1988M}, $\sim41$\,mHz QPO in the transient high-mass Be/X-ray binary pulsar 4U 0115+634 using data from the Rossi X-Ray Timing Explorer~\citep{2013MNRAS.434.2458D}, the physical mechanism of these high energy mHz QPOs is still needed to be revealed~\citep{2000ApJ...545..399B,2009A&A...493..809B}.

To understand the X-ray burst phenomenon as well as marginally stable burning from accreting SS, one needs to focus the microphysics inside SS and the nuclear reactions included rp processes in the envelope, and perform numerical calculations of the thermal evolution of SSs using a general relativistic stellar-evolution code~\citep{2020PTEP.2020c3E02D,2021ApJ...923...64D,2022ApJ...937..124D}. Such a detailed study is left for our future work.

\section*{Acknowledgements}

We thank the referee for the constructive feedback and comments that greatly improved the quality of this paper. We thank M. Hashimoto for his encouragement. This work received the generous support of the Major Science and Technology Program of Xinjiang Uygur Autonomous Region under Grant No. 2022A03013-3, the National Natural Science Foundation of China Nos. 12263006, U2031204, 12273030, 12288102.

\section*{Data Availability}
The data underlying this article are available in this article.



\bibliographystyle{mnras}
\bibliography{my} 

\begin{thebibliography}{}
\makeatletter
\relax
\def\mn@urlcharsother{\let\do\@makeother \do\$\do\&\do\#\do\^\do\_\do\%\do\~}
\def\mn@doi{\begingroup\mn@urlcharsother \@ifnextchar [ {\mn@doi@}
  {\mn@doi@[]}}
\def\mn@doi@[#1]#2{\def\@tempa{#1}\ifx\@tempa\@empty \href
  {http://dx.doi.org/#2} {doi:#2}\else \href {http://dx.doi.org/#2} {#1}\fi
  \endgroup}
\def\mn@eprint#1#2{\mn@eprint@#1:#2::\@nil}
\def\mn@eprint@arXiv#1{\href {http://arxiv.org/abs/#1} {{\tt arXiv:#1}}}
\def\mn@eprint@dblp#1{\href {http://dblp.uni-trier.de/rec/bibtex/#1.xml}
  {dblp:#1}}
\def\mn@eprint@#1:#2:#3:#4\@nil{\def\@tempa {#1}\def\@tempb {#2}\def\@tempc
  {#3}\ifx \@tempc \@empty \let \@tempc \@tempb \let \@tempb \@tempa \fi \ifx
  \@tempb \@empty \def\@tempb {arXiv}\fi \@ifundefined
  {mn@eprint@\@tempb}{\@tempb:\@tempc}{\expandafter \expandafter \csname
  mn@eprint@\@tempb\endcsname \expandafter{\@tempc}}}

\bibitem[\protect\citeauthoryear{{Alcock}, {Farhi}  \& {Olinto}}{{Alcock}
  et~al.}{1986}]{1986ApJ...310..261A}
{Alcock} C.,  {Farhi} E.,   {Olinto} A.,  1986, \mn@doi [\apj]
  {10.1086/164679}, \href
  {https://ui.adsabs.harvard.edu/abs/1986ApJ...310..261A} {310, 261}

\bibitem[\protect\citeauthoryear{{Altamirano}, {van der Klis}, {Wijnands}  \&
  {Cumming}}{{Altamirano} et~al.}{2008}]{2008ApJ...673L..35A}
{Altamirano} D.,  {van der Klis} M.,  {Wijnands} R.,   {Cumming} A.,  2008,
  \mn@doi [\apjl] {10.1086/527355}, \href
  {https://ui.adsabs.harvard.edu/abs/2008ApJ...673L..35A} {673, L35}

\bibitem[\protect\citeauthoryear{{Antoniadis} et~al.,}{{Antoniadis}
  et~al.}{2013}]{2013Sci...340..448A}
{Antoniadis} J.,  et~al., 2013, \mn@doi [Science] {10.1126/science.1233232},
  \href {https://ui.adsabs.harvard.edu/abs/2013Sci...340..448A} {340, 448}

\bibitem[\protect\citeauthoryear{{Bildsten}}{{Bildsten}}{1998}]{1998ASIC..515..419B}
{Bildsten} L.,  1998, in {Buccheri} R.,  {van Paradijs} J.,   {Alpar} A.,  eds,
   NATO Advanced Study Institute (ASI) Series C Vol. 515, The Many Faces of
  Neutron Stars.. p.~419 (\mn@eprint {arXiv} {astro-ph/9709094}),
  \mn@doi{10.48550/arXiv.astro-ph/9709094}

\bibitem[\protect\citeauthoryear{{Boroson}, {O'Brien}, {Horne}, {Kallman},
  {Still}, {Boyd}, {Quaintrell}  \& {Vrtilek}}{{Boroson}
  et~al.}{2000}]{2000ApJ...545..399B}
{Boroson} B.,  {O'Brien} K.,  {Horne} K.,  {Kallman} T.,  {Still} M.,  {Boyd}
  P.~T.,  {Quaintrell} H.,   {Vrtilek} S.~D.,  2000, \mn@doi [\apj]
  {10.1086/317786}, \href
  {https://ui.adsabs.harvard.edu/abs/2000ApJ...545..399B} {545, 399}

\bibitem[\protect\citeauthoryear{{Bozzo}, {Stella}, {Vietri}  \&
  {Ghosh}}{{Bozzo} et~al.}{2009}]{2009A&A...493..809B}
{Bozzo} E.,  {Stella} L.,  {Vietri} M.,   {Ghosh} P.,  2009, \mn@doi [\aap]
  {10.1051/0004-6361:200810658}, \href
  {https://ui.adsabs.harvard.edu/abs/2009A&A...493..809B} {493, 809}

\bibitem[\protect\citeauthoryear{{Cumming}, {Macbeth}, {in 't Zand}  \&
  {Page}}{{Cumming} et~al.}{2006}]{2006ApJ...646..429C}
{Cumming} A.,  {Macbeth} J.,  {in 't Zand} J.~J.~M.,   {Page} D.,  2006,
  \mn@doi [\apj] {10.1086/504698}, \href
  {https://ui.adsabs.harvard.edu/abs/2006ApJ...646..429C} {646, 429}

\bibitem[\protect\citeauthoryear{{Deibel}, {Cumming}, {Brown}  \&
  {Page}}{{Deibel} et~al.}{2015}]{2015ApJ...809L..31D}
{Deibel} A.,  {Cumming} A.,  {Brown} E.~F.,   {Page} D.,  2015, \mn@doi [\apjl]
  {10.1088/2041-8205/809/2/L31}, \href
  {https://ui.adsabs.harvard.edu/abs/2015ApJ...809L..31D} {809, L31}

\bibitem[\protect\citeauthoryear{{Demorest}, {Pennucci}, {Ransom}, {Roberts}
  \& {Hessels}}{{Demorest} et~al.}{2010}]{2010Natur.467.1081D}
{Demorest} P.~B.,  {Pennucci} T.,  {Ransom} S.~M.,  {Roberts} M.~S.~E.,
  {Hessels} J.~W.~T.,  2010, \mn@doi [\nat] {10.1038/nature09466}, \href
  {https://ui.adsabs.harvard.edu/abs/2010Natur.467.1081D} {467, 1081}

\bibitem[\protect\citeauthoryear{{Dohi}, {Hashimoto}, {Yamada}, {Matsuo}  \&
  {Fujimoto}}{{Dohi} et~al.}{2020}]{2020PTEP.2020c3E02D}
{Dohi} A.,  {Hashimoto} M.-a.,  {Yamada} R.,  {Matsuo} Y.,   {Fujimoto} M.~Y.,
  2020, \mn@doi [Progress of Theoretical and Experimental Physics]
  {10.1093/ptep/ptaa010}, \href
  {https://ui.adsabs.harvard.edu/abs/2020PTEP.2020c3E02D} {2020, 033E02}

\bibitem[\protect\citeauthoryear{{Dohi}, {Nishimura}, {Hashimoto}, {Matsuo},
  {Noda}  \& {Nagataki}}{{Dohi} et~al.}{2021}]{2021ApJ...923...64D}
{Dohi} A.,  {Nishimura} N.,  {Hashimoto} M.,  {Matsuo} Y.,  {Noda} T.,
  {Nagataki} S.,  2021, \apj, \href
  {https://ui.adsabs.harvard.edu/abs/2021ApJ...923...64D} {923, 64}

\bibitem[\protect\citeauthoryear{{Dohi}, {Nishimura}, {Sotani}, {Noda}, {Liu},
  {Nagataki}  \& {Hashimoto}}{{Dohi} et~al.}{2022}]{2022ApJ...937..124D}
{Dohi} A.,  {Nishimura} N.,  {Sotani} H.,  {Noda} T.,  {Liu} H.-L.,  {Nagataki}
  S.,   {Hashimoto} M.,  2022, \mn@doi [\apj] {10.3847/1538-4357/ac8dfe}, \href
  {https://ui.adsabs.harvard.edu/abs/2022ApJ...937..124D} {937, 124}

\bibitem[\protect\citeauthoryear{{Dugair}, {Jaisawal}, {Naik}  \&
  {Jaaffrey}}{{Dugair} et~al.}{2013}]{2013MNRAS.434.2458D}
{Dugair} M.~R.,  {Jaisawal} G.~K.,  {Naik} S.,   {Jaaffrey} S.~N.~A.,  2013,
  \mn@doi [\mnras] {10.1093/mnras/stt1187}, \href
  {https://ui.adsabs.harvard.edu/abs/2013MNRAS.434.2458D} {434, 2458}

\bibitem[\protect\citeauthoryear{{Farhi} \& {Jaffe}}{{Farhi} \&
  {Jaffe}}{1984}]{1984PhRvD..30.2379F}
{Farhi} E.,  {Jaffe} R.~L.,  1984, \mn@doi [\prd] {10.1103/PhysRevD.30.2379},
  \href {https://ui.adsabs.harvard.edu/abs/1984PhRvD..30.2379F} {30, 2379}

\bibitem[\protect\citeauthoryear{{Fattoyev}, {Brown}, {Cumming}, {Deibel},
  {Horowitz}, {Li}  \& {Lin}}{{Fattoyev} et~al.}{2018}]{2018PhRvC..98b5801F}
{Fattoyev} F.~J.,  {Brown} E.~F.,  {Cumming} A.,  {Deibel} A.,  {Horowitz}
  C.~J.,  {Li} B.-A.,   {Lin} Z.,  2018, \mn@doi [\prc]
  {10.1103/PhysRevC.98.025801}, \href
  {https://ui.adsabs.harvard.edu/abs/2018PhRvC..98b5801F} {98, 025801}

\bibitem[\protect\citeauthoryear{{Ferrigno} et~al.,}{{Ferrigno}
  et~al.}{2017}]{2017MNRAS.466.3450F}
{Ferrigno} C.,  et~al., 2017, \mn@doi [\mnras] {10.1093/mnras/stw3344}, \href
  {https://ui.adsabs.harvard.edu/abs/2017MNRAS.466.3450F} {466, 3450}

\bibitem[\protect\citeauthoryear{{Fonseca} et~al.,}{{Fonseca}
  et~al.}{2021}]{2021ApJ...915L..12F}
{Fonseca} E.,  et~al., 2021, \mn@doi [\apjl] {10.3847/2041-8213/ac03b8}, \href
  {https://ui.adsabs.harvard.edu/abs/2021ApJ...915L..12F} {915, L12}

\bibitem[\protect\citeauthoryear{{Fowler}}{{Fowler}}{1926}]{Fowler1926}
{Fowler} R.~H.,  1926, \mn@doi [\mnras] {10.1093/mnras/87.2.114}, \href
  {https://ui.adsabs.harvard.edu/abs/1926MNRAS..87..114F} {87, 114}

\bibitem[\protect\citeauthoryear{{Fujimoto}, {Hanawa}  \& {Miyaji}}{{Fujimoto}
  et~al.}{1981}]{1981ApJ...247..267F}
{Fujimoto} M.~Y.,  {Hanawa} T.,   {Miyaji} S.,  1981, \mn@doi [\apj]
  {10.1086/159034}, \href
  {https://ui.adsabs.harvard.edu/abs/1981ApJ...247..267F} {247, 267}

\bibitem[\protect\citeauthoryear{{Fushiki} \& {Lamb}}{{Fushiki} \&
  {Lamb}}{1987}]{1987ApJ...317..368F}
{Fushiki} I.,  {Lamb} D.~Q.,  1987, \mn@doi [\apj] {10.1086/165284}, \href
  {https://ui.adsabs.harvard.edu/abs/1987ApJ...317..368F} {317, 368}

\bibitem[\protect\citeauthoryear{{Galloway} \& {Keek}}{{Galloway} \&
  {Keek}}{2021}]{2021ASSL..461..209G}
{Galloway} D.~K.,  {Keek} L.,  2021, in {Belloni} T.~M.,  {M{\'e}ndez} M.,
  {Zhang} C.,  eds,  Astrophysics and Space Science Library Vol. 461, Timing
  Neutron Stars: Pulsations, Oscillations and Explosions. pp 209--262
  (\mn@eprint {arXiv} {1712.06227}), \mn@doi{10.1007/978-3-662-62110-3_5}

\bibitem[\protect\citeauthoryear{{Gao}, {Lai}, {Shao}  \& {Xu}}{{Gao}
  et~al.}{2022}]{2022MNRAS.509.2758G}
{Gao} Y.,  {Lai} X.-Y.,  {Shao} L.,   {Xu} R.-X.,  2022, \mn@doi [\mnras]
  {10.1093/mnras/stab3181}, \href
  {https://ui.adsabs.harvard.edu/abs/2022MNRAS.509.2758G} {509, 2758}

\bibitem[\protect\citeauthoryear{{Gupta}, {Brown}, {Schatz}, {M{\"o}ller}  \&
  {Kratz}}{{Gupta} et~al.}{2007}]{2007ApJ...662.1188G}
{Gupta} S.,  {Brown} E.~F.,  {Schatz} H.,  {M{\"o}ller} P.,   {Kratz} K.-L.,
  2007, \mn@doi [\apj] {10.1086/517869}, \href
  {https://ui.adsabs.harvard.edu/abs/2007ApJ...662.1188G} {662, 1188}

\bibitem[\protect\citeauthoryear{{Haensel} \& {Zdunik}}{{Haensel} \&
  {Zdunik}}{1990}]{1990A&A...227..431H}
{Haensel} P.,  {Zdunik} J.~L.,  1990, \aap, \href
  {https://ui.adsabs.harvard.edu/abs/1990A&A...227..431H} {227, 431}

\bibitem[\protect\citeauthoryear{{Haensel} \& {Zdunik}}{{Haensel} \&
  {Zdunik}}{1991}]{1991NuPhS..24..139H}
{Haensel} P.,  {Zdunik} J.~L.,  1991, \mn@doi [Nuclear Physics B Proceedings
  Supplements] {10.1016/0920-5632(91)90312-3}, \href
  {https://ui.adsabs.harvard.edu/abs/1991NuPhS..24..139H} {24, 139}

\bibitem[\protect\citeauthoryear{{Haensel} \& {Zdunik}}{{Haensel} \&
  {Zdunik}}{2003}]{2003A&A...404L..33H}
{Haensel} P.,  {Zdunik} J.~L.,  2003, \mn@doi [\aap]
  {10.1051/0004-6361:20030708}, \href
  {https://ui.adsabs.harvard.edu/abs/2003A&A...404L..33H} {404, L33}

\bibitem[\protect\citeauthoryear{{Haensel} \& {Zdunik}}{{Haensel} \&
  {Zdunik}}{2008}]{2008A&A...480..459H}
{Haensel} P.,  {Zdunik} J.~L.,  2008, \mn@doi [\aap]
  {10.1051/0004-6361:20078578}, \href
  {https://ui.adsabs.harvard.edu/abs/2008A&A...480..459H} {480, 459}

\bibitem[\protect\citeauthoryear{{Haensel}, {Zdunik}  \& {Schaeffer}}{{Haensel}
  et~al.}{1986}]{Haensel1986}
{Haensel} P.,  {Zdunik} J.~L.,   {Schaeffer} R.,  1986, \aap, \href
  {https://ui.adsabs.harvard.edu/abs/1986A&A...160..251H} {160, 251}

\bibitem[\protect\citeauthoryear{{Heger}, {Cumming}  \& {Woosley}}{{Heger}
  et~al.}{2007}]{2007ApJ...665.1311H}
{Heger} A.,  {Cumming} A.,   {Woosley} S.~E.,  2007, \mn@doi [\apj]
  {10.1086/517491}, \href
  {https://ui.adsabs.harvard.edu/abs/2007ApJ...665.1311H} {665, 1311}

\bibitem[\protect\citeauthoryear{{Johnston}, {Heger}  \& {Galloway}}{{Johnston}
  et~al.}{2020}]{2020MNRAS.494.4576J}
{Johnston} Z.,  {Heger} A.,   {Galloway} D.~K.,  2020, \mn@doi [\mnras]
  {10.1093/mnras/staa1054}, \href
  {https://ui.adsabs.harvard.edu/abs/2020MNRAS.494.4576J} {494, 4576}

\bibitem[\protect\citeauthoryear{{Keek}, {Langer}  \& {in't Zand}}{{Keek}
  et~al.}{2009}]{2009A&A...502..871K}
{Keek} L.,  {Langer} N.,   {in't Zand} J.~J.~M.,  2009, \mn@doi [\aap]
  {10.1051/0004-6361/200911619}, \href
  {https://ui.adsabs.harvard.edu/abs/2009A&A...502..871K} {502, 871}

\bibitem[\protect\citeauthoryear{{Keek}, {Cyburt}  \& {Heger}}{{Keek}
  et~al.}{2014}]{2014ApJ...787..101K}
{Keek} L.,  {Cyburt} R.~H.,   {Heger} A.,  2014, \mn@doi [\apj]
  {10.1088/0004-637X/787/2/101}, \href
  {https://ui.adsabs.harvard.edu/abs/2014ApJ...787..101K} {787, 101}

\bibitem[\protect\citeauthoryear{{Lai} \& {Xu}}{{Lai} \&
  {Xu}}{2009}]{2009MNRAS.398L..31L}
{Lai} X.~Y.,  {Xu} R.~X.,  2009, \mn@doi [\mnras]
  {10.1111/j.1745-3933.2009.00701.x}, \href
  {https://ui.adsabs.harvard.edu/abs/2009MNRAS.398L..31L} {398, L31}

\bibitem[\protect\citeauthoryear{{Landau}}{{Landau}}{1932}]{Landau1932}
{Landau} L.~D.,  1932, Phys. Zs. Sowjet, \href
  {https://ui.adsabs.harvard.edu/abs/1932PhyZS...1..285L} {1, 285}

\bibitem[\protect\citeauthoryear{{Lewin}, {van Paradijs}  \& {Taam}}{{Lewin}
  et~al.}{1993}]{1993SSRv...62..223L}
{Lewin} W. H.~G.,  {van Paradijs} J.,   {Taam} R.~E.,  1993, \mn@doi [\ssr]
  {10.1007/BF00196124}, \href
  {https://ui.adsabs.harvard.edu/abs/1993SSRv...62..223L} {62, 223}

\bibitem[\protect\citeauthoryear{{Li}, {Pan}  \& {Falanga}}{{Li}
  et~al.}{2021}]{2021ApJ...920...35L}
{Li} Z.,  {Pan} Y.,   {Falanga} M.,  2021, \mn@doi [\apj]
  {10.3847/1538-4357/ac1f15}, \href
  {https://ui.adsabs.harvard.edu/abs/2021ApJ...920...35L} {920, 35}

\bibitem[\protect\citeauthoryear{{Linares} et~al.,}{{Linares}
  et~al.}{2010}]{2010ATel.2958....1L}
{Linares} M.,  et~al., 2010, The Astronomer's Telegram, \href
  {https://ui.adsabs.harvard.edu/abs/2010ATel.2958....1L} {2958, 1}

\bibitem[\protect\citeauthoryear{{Liu}, {Dai}, {L{\"u}}, {Dohi}, {Yong}  \&
  {Hashimoto}}{{Liu} et~al.}{2021}]{2021PhRvD.104l3004L}
{Liu} H.-L.,  {Dai} Z.-G.,  {L{\"u}} G.-L.,  {Dohi} A.,  {Yong} G.-C.,
  {Hashimoto} M.-a.,  2021, \mn@doi [\prd] {10.1103/PhysRevD.104.123004}, \href
  {https://ui.adsabs.harvard.edu/abs/2021PhRvD.104l3004L} {104, 123004}

\bibitem[\protect\citeauthoryear{{Lyu}, {M{\'e}ndez}, {Zhang}  \& {Keek}}{{Lyu}
  et~al.}{2015}]{2015MNRAS.454..541L}
{Lyu} M.,  {M{\'e}ndez} M.,  {Zhang} G.,   {Keek} L.,  2015, \mn@doi [\mnras]
  {10.1093/mnras/stv1971}, \href
  {https://ui.adsabs.harvard.edu/abs/2015MNRAS.454..541L} {454, 541}

\bibitem[\protect\citeauthoryear{{Ma} et~al.,}{{Ma}
  et~al.}{2022}]{2022MNRAS.517.1988M}
{Ma} R.,  et~al., 2022, \mn@doi [\mnras] {10.1093/mnras/stac2768}, \href
  {https://ui.adsabs.harvard.edu/abs/2022MNRAS.517.1988M} {517, 1988}

\bibitem[\protect\citeauthoryear{{Mancuso}, {Altamirano}, {Garc{\'\i}a}, {Lyu},
  {M{\'e}ndez}, {Combi}, {D{\'\i}az-Trigo}  \& {in't Zand}}{{Mancuso}
  et~al.}{2019}]{2019MNRAS.486L..74M}
{Mancuso} G.~C.,  {Altamirano} D.,  {Garc{\'\i}a} F.,  {Lyu} M.,  {M{\'e}ndez}
  M.,  {Combi} J.~A.,  {D{\'\i}az-Trigo} M.,   {in't Zand} J.~J.~M.,  2019,
  \mn@doi [\mnras] {10.1093/mnrasl/slz057}, \href
  {https://ui.adsabs.harvard.edu/abs/2019MNRAS.486L..74M} {486, L74}

\bibitem[\protect\citeauthoryear{{Mancuso} et~al.,}{{Mancuso}
  et~al.}{2023}]{2023MNRAS.521.5616M}
{Mancuso} G.~C.,  et~al., 2023, \mn@doi [\mnras] {10.1093/mnras/stad949}, \href
  {https://ui.adsabs.harvard.edu/abs/2023MNRAS.521.5616M} {521, 5616}

\bibitem[\protect\citeauthoryear{{Meisel}}{{Meisel}}{2018}]{2018ApJ...860..147M}
{Meisel} Z.,  2018, \mn@doi [\apj] {10.3847/1538-4357/aac3d3}, \href
  {https://ui.adsabs.harvard.edu/abs/2018ApJ...860..147M} {860, 147}

\bibitem[\protect\citeauthoryear{{Paczynski}}{{Paczynski}}{1983}]{1983ApJ...264..282P}
{Paczynski} B.,  1983, \mn@doi [\apj] {10.1086/160596}, \href
  {https://ui.adsabs.harvard.edu/abs/1983ApJ...264..282P} {264, 282}

\bibitem[\protect\citeauthoryear{{Page} \& {Cumming}}{{Page} \&
  {Cumming}}{2005}]{2005ApJ...635L.157P}
{Page} D.,  {Cumming} A.,  2005, \mn@doi [\apjl] {10.1086/499520}, \href
  {https://ui.adsabs.harvard.edu/abs/2005ApJ...635L.157P} {635, L157}

\bibitem[\protect\citeauthoryear{{Piro} \& {Bildsten}}{{Piro} \&
  {Bildsten}}{2007}]{2007ApJ...663.1252P}
{Piro} A.~L.,  {Bildsten} L.,  2007, \mn@doi [\apj] {10.1086/518687}, \href
  {https://ui.adsabs.harvard.edu/abs/2007ApJ...663.1252P} {663, 1252}

\bibitem[\protect\citeauthoryear{{Prakash}, {Lattimer}, {Steiner}  \&
  {Page}}{{Prakash} et~al.}{2003}]{2003NuPhA.715..835P}
{Prakash} M.,  {Lattimer} J.~M.,  {Steiner} A.~W.,   {Page} D.,  2003, \mn@doi
  [\nphysa] {10.1016/S0375-9474(02)01514-2}, \href
  {https://ui.adsabs.harvard.edu/abs/2003NuPhA.715..835P} {715, 835}

\bibitem[\protect\citeauthoryear{{Revnivtsev}, {Churazov}, {Gilfanov}  \&
  {Sunyaev}}{{Revnivtsev} et~al.}{2001}]{2001A&A...372..138R}
{Revnivtsev} M.,  {Churazov} E.,  {Gilfanov} M.,   {Sunyaev} R.,  2001, \mn@doi
  [\aap] {10.1051/0004-6361:20010434}, \href
  {https://ui.adsabs.harvard.edu/abs/2001A&A...372..138R} {372, 138}

\bibitem[\protect\citeauthoryear{{Schatz}, {Bildsten}, {Cumming}  \&
  {Wiescher}}{{Schatz} et~al.}{1999}]{1999ApJ...524.1014S}
{Schatz} H.,  {Bildsten} L.,  {Cumming} A.,   {Wiescher} M.,  1999, \mn@doi
  [\apj] {10.1086/307837}, \href
  {https://ui.adsabs.harvard.edu/abs/1999ApJ...524.1014S} {524, 1014}

\bibitem[\protect\citeauthoryear{{Strohmayer} \& {Bildsten}}{{Strohmayer} \&
  {Bildsten}}{2006}]{2006csxs.book..113S}
{Strohmayer} T.,  {Bildsten} L.,  2006, in , Vol.~39, Compact stellar X-ray
  sources.
pp 113--156

\bibitem[\protect\citeauthoryear{{Strohmayer} \& {Smith}}{{Strohmayer} \&
  {Smith}}{2011}]{2011ATel.3258....1S}
{Strohmayer} T.~E.,  {Smith} E.~A.,  2011, The Astronomer's Telegram, \href
  {https://ui.adsabs.harvard.edu/abs/2011ATel.3258....1S} {3258, 1}

\bibitem[\protect\citeauthoryear{{Strohmayer} et~al.,}{{Strohmayer}
  et~al.}{2018}]{2018ApJ...865...63S}
{Strohmayer} T.~E.,  et~al., 2018, \mn@doi [\apj] {10.3847/1538-4357/aada14},
  \href {https://ui.adsabs.harvard.edu/abs/2018ApJ...865...63S} {865, 63}

\bibitem[\protect\citeauthoryear{{Togashi}, {Nakazato}, {Takehara}, {Yamamuro},
  {Suzuki}  \& {Takano}}{{Togashi} et~al.}{2017}]{2017NuPhA.961...78T}
{Togashi} H.,  {Nakazato} K.,  {Takehara} Y.,  {Yamamuro} S.,  {Suzuki} H.,
  {Takano} M.,  2017, \mn@doi [\nphysa] {10.1016/j.nuclphysa.2017.02.010},
  \href {https://ui.adsabs.harvard.edu/abs/2017NuPhA.961...78T} {961, 78}

\bibitem[\protect\citeauthoryear{{Tse}, {Galloway}, {Chou}, {Heger}  \&
  {Hsieh}}{{Tse} et~al.}{2021}]{2021MNRAS.500...34T}
{Tse} K.,  {Galloway} D.~K.,  {Chou} Y.,  {Heger} A.,   {Hsieh} H.-E.,  2021,
  \mn@doi [\mnras] {10.1093/mnras/staa3224}, \href
  {https://ui.adsabs.harvard.edu/abs/2021MNRAS.500...34T} {500, 34}

\bibitem[\protect\citeauthoryear{{Wallace} \& {Woosley}}{{Wallace} \&
  {Woosley}}{1981}]{1981ApJS...45..389W}
{Wallace} R.~K.,  {Woosley} S.~E.,  1981, \mn@doi [\apjs] {10.1086/190717},
  \href {https://ui.adsabs.harvard.edu/abs/1981ApJS...45..389W} {45, 389}

\bibitem[\protect\citeauthoryear{{Witten}}{{Witten}}{1984}]{1984PhRvD..30..272W}
{Witten} E.,  1984, \mn@doi [\prd] {10.1103/PhysRevD.30.272}, \href
  {https://ui.adsabs.harvard.edu/abs/1984PhRvD..30..272W} {30, 272}

\bibitem[\protect\citeauthoryear{{Woosley} et~al.,}{{Woosley}
  et~al.}{2004}]{2004ApJS..151...75W}
{Woosley} S.~E.,  et~al., 2004, \mn@doi [\apjs] {10.1086/381533}, \href
  {https://ui.adsabs.harvard.edu/abs/2004ApJS..151...75W} {151, 75}

\bibitem[\protect\citeauthoryear{{Xiaoyu} \& {Renxin}}{{Xiaoyu} \&
  {Renxin}}{2017}]{2017JPhCS.861a2027X}
{Xiaoyu} L.,  {Renxin} X.,  2017, in Journal of Physics Conference Series. p.
  012027 (\mn@eprint {arXiv} {1701.08463}),
  \mn@doi{10.1088/1742-6596/861/1/012027}

\bibitem[\protect\citeauthoryear{{Xu}}{{Xu}}{2003}]{2003ApJ...596L..59X}
{Xu} R.~X.,  2003, \mn@doi [\apjl] {10.1086/379209}, \href
  {https://ui.adsabs.harvard.edu/abs/2003ApJ...596L..59X} {596, L59}

\bibitem[\protect\citeauthoryear{{Xu}}{{Xu}}{2023}]{Xu2023}
{Xu} R.,  2023, \mn@doi [Astron. Nachr.] {10.1002/asna.20230008}, \href
  {https://ui.adsabs.harvard.edu/abs/2022arXiv221210887X} {p. e20230008
  (arXiv:2212.10887)}

\bibitem[\protect\citeauthoryear{{Yakovlev}, {Haensel}, {Baym}  \&
  {Pethick}}{{Yakovlev} et~al.}{2013}]{Yakovlev2013}
{Yakovlev} D.~G.,  {Haensel} P.,  {Baym} G.,   {Pethick} C.,  2013, \mn@doi
  [Physics Uspekhi] {10.3367/UFNe.0183.201303f.0307}, \href
  {https://ui.adsabs.harvard.edu/abs/2013PhyU...56..289Y} {56, 289}

\bibitem[\protect\citeauthoryear{{Zhang}, {Geng}  \& {Huang}}{{Zhang}
  et~al.}{2018}]{2018ApJ...858...88Z}
{Zhang} Y.,  {Geng} J.-J.,   {Huang} Y.-F.,  2018, \mn@doi [\apj]
  {10.3847/1538-4357/aabaee}, \href
  {https://ui.adsabs.harvard.edu/abs/2018ApJ...858...88Z} {858, 88}

\bibitem[\protect\citeauthoryear{{Zhen} et~al.,}{{Zhen}
  et~al.}{2023}]{2023ApJ...950..110Z}
{Zhen} G.,  et~al., 2023, \mn@doi [\apj] {10.3847/1538-4357/accd5f}, \href
  {https://ui.adsabs.harvard.edu/abs/2023ApJ...950..110Z} {950, 110}

\bibitem[\protect\citeauthoryear{{Zhu}, {L{\"u}}, {Wang}  \& {Liu}}{{Zhu}
  et~al.}{2013}]{2013PASP..125...25Z}
{Zhu} C.,  {L{\"u}} G.,  {Wang} Z.,   {Liu} J.,  2013, \mn@doi [\pasp]
  {10.1086/669193}, \href
  {https://ui.adsabs.harvard.edu/abs/2013PASP..125...25Z} {125, 25}

\bibitem[\protect\citeauthoryear{{in't Zand} et~al.,}{{in't Zand}
  et~al.}{2004}]{2004NuPhS.132..486I}
{in't Zand} J.,  et~al., 2004, \mn@doi [Nuclear Physics B Proceedings
  Supplements] {10.1016/j.nuclphysbps.2004.04.083}, \href
  {https://ui.adsabs.harvard.edu/abs/2004NuPhS.132..486I} {132, 486}

\makeatother
\end{thebibliography}




\appendix
\begin{table*}
    \centering
    \caption{Physical quantities of oscillation frequency $f$, critical mass accretion rate $\dot{m}/\dot{m}_{\rm Edd}$ (the last value of the critical mass accretion rate in its narrow range), maxima luminosity $L_{36}(\rm max)$, minima luminosity $L_{36}(\rm min)$ with variation in surface gravity $g_{14}$, base heating $Q_{\rm b}$, helium mass fraction $Y_0$, where the surface gravities are inferred from Togashi EoS, $L_{36}={L}/{10^{36}}\,{\rm erg\,s^{-1}}$, the metallicity is set as $Z=0.01$. The "-" in the tables means that it is in stable burning when $Q_{\rm b}$ is high enough. $f_{\infty}$ represents the general relativity correction for frequency.
    }
    \begin{tabular}{ccccccccccc}
    \toprule
    $Y_0=0.3$ & \multicolumn{5}{c}{$g_{14}=0.50 ~(0.5\,M_{\odot})$}& \multicolumn{5}{c}{$g_{14}=1.01~(1.0\,M_{\odot})$}\\
    \cmidrule(r){2-6}\cmidrule(r){7-11}
    $Q_{\rm b}(\rm MeV/u)$ & $f(\rm mHz)$ & $f_{\infty}(\rm mHz)$ & $\dot{m}/\dot{m}_{\rm Edd}$ & $L_{36}(\rm max)$ & $L_{36}(\rm min)$
      & $f(\rm mHz)$ & $f_{\infty}(\rm mHz)$ & $\dot{m}/\dot{m}_{\rm Edd}$  & $L_{36}(\rm max)$ & $L_{36}(\rm min)$\\
        \midrule
0 & 0.75 & 0.70 & 0.402 & 4.12 & 2.16 & 1.71 & 1.50 & 0.667 & 6.97 & 3.52\\
0.15 & 0.70 & 0.65 & 0.380 & 4.53 & 1.79 & 1.63 & 1.43 & 0.641 & 6.59 & 3.44\\ 
0.5 & 0.68 & 0.64 & 0.341 & 3.45 & 1.87 & 1.57 & 1.38 & 0.577 & 5.87 & 3.16\\
1.0 & 0.55 & 0.51 & 0.273 & 2.69 & 1.55 & 1.34 & 1.18 & 0.482 & 4.71 & 2.74\\
1.5 & 0.39 & 0.36 & 0.202 & 1.92 & 1.18 & 0.93 & 0.82 & 0.382 & 3.67 & 2.20\\
2.0 & - & - & - & - & - & 0.73 & 0.64 & 0.274 & 2.69 & 1.57\\

\toprule
    \multirow{2}{*}{} & \multicolumn{5}{c}{$g_{14}=1.50 ~(1.5\,M_{\odot})$}& \multicolumn{5}{c}{$g_{14}=2.12~(2.0\,M_{\odot})$}\\
    \cmidrule(r){2-6}\cmidrule(r){7-11}
     & $f(\rm mHz)$ & $f_{\infty}(\rm mHz)$ & $\dot{m}/\dot{m}_{\rm Edd}$ & $L_{36}(\rm max)$ & $L_{36}(\rm min)$
      & $f(\rm mHz)$ & $f_{\infty}(\rm mHz)$  & $\dot{m}/\dot{m}_{\rm Edd}$  & $L_{36}(\rm max)$ & $L_{36}(\rm min)$\\
        \midrule
0 & 2.70 & 2.23 & 0.879 & 8.85 & 4.89 & 3.85 & 2.96 & 1.125 & 11.17 & 5.62\\
0.15 & 2.53 & 2.09 & 0.85 & 8.76 & 4.66 & 3.80 & 2.92 & 1.081 & 10.80 & 5.37\\
0.5 & 2.41 & 1.99 & 0.760 & 7.98 & 4.14 & 3.36 & 2.58 & 0.975 & 9.90 & 4.81\\
1.0 & 1.95 & 1.61 & 0.642 & 6.68 & 3.50 & 2.85 & 2.19 & 0.820 & 8.56 & 4.00\\
1.5 & 1.56 & 1.29 & 0.516 & 5.20 & 2.91 & 2.53 & 1.95 & 0.67 & 6.32 & 3.58\\
2.0 & 1.19 & 0.98 & 0.385 & 3.85 & 2.22 & 1.80 & 1.38 & 0.508 & 4.68 & 2.77\\

\toprule
    $Y_0=0.4$ & \multicolumn{5}{c}{$g_{14}=0.50 ~(0.5\,M_{\odot})$}& \multicolumn{5}{c}{$g_{14}=1.01~(1.0\,M_{\odot})$}\\
    \cmidrule(r){2-6}\cmidrule(r){7-11}
    $Q_{\rm b}(\rm MeV/u)$ & $f(\rm mHz)$ & $f_{\infty}(\rm mHz)$ & $\dot{m}/\dot{m}_{\rm Edd}$ & $L_{36}(\rm max)$ & $L_{36}(\rm min)$
      & $f(\rm mHz)$ & $f_{\infty}(\rm mHz)$ & $\dot{m}/\dot{m}_{\rm Edd}$  & $L_{36}(\rm max)$ & $L_{36}(\rm min)$\\
        \midrule
0 & 1.27 & 1.19 & 0.517 & 5.30 & 2.76 & 2.91 & 2.55 & 0.842 & 9.07 & 4.31\\
0.15 & 1.31 & 1.22 & 0.496 & 4.96 & 2.73 & 2.76 & 2.42 & 0.812 & 8.28 & 4.38\\
0.5 & 1.19 & 1.11 & 0.443 & 4.44 & 2.43 & 2.62 & 2.30 & 0.732 & 7.47 & 3.96\\
1.0 & 0.99 & 0.93 & 0.365 & 3.61 & 2.04 & 2.10 & 1.84 & 0.613 & 6.26 & 3.33\\
1.5 & 0.74 & 0.69 & 0.281 & 2.75 & 1.60 & 1.69 & 1.48 & 0.491 & 4.92 & 2.75\\
2.0 & - & - & - & - & - & 1.38 & 1.21 & 0.368 & 3.55 & 2.14\\
\toprule
    \multirow{2}{*}{} & \multicolumn{5}{c}{$g_{14}=1.50 ~(1.5\,M_{\odot})$}& \multicolumn{5}{c}{$g_{14}=2.12~(2.0\,M_{\odot})$}\\
    \cmidrule(r){2-6}\cmidrule(r){7-11}
     & $f(\rm mHz)$ & $f_{\infty}(\rm mHz)$ & $\dot{m}/\dot{m}_{\rm Edd}$ & $L_{36}(\rm max)$ & $L_{36}(\rm min)$
      & $f(\rm mHz)$ & $f_{\infty}(\rm mHz)$ & $\dot{m}/\dot{m}_{\rm Edd}$  & $L_{36}(\rm max)$ & $L_{36}(\rm min)$\\
        \midrule
0 & 4.55 & 3.76 & 1.109 & 11.66 & 5.94 & 6.56 & 5.05 & 1.41 & 13.86 & 7.07 \\
0.15 & 4.17 & 3.45 & 1.065 & 11.23 & 5.70 & 6.46 & 4.97 & 1.355 & 13.52 & 6.73\\ 
0.5 & 3.89 & 3.21 & 0.961 & 10.51 & 5.07 & 5.75 & 4.42 & 1.225 & 12.27 & 6.13\\
1.0 & 3.49 & 2.88 & 0.815 & 8.39 & 4.55 & 4.83 & 3.72 & 1.025 & 11.13 & 4.83\\
1.5 & 2.80 & 2.31 & 0.651 & 7.03 & 3.48 & 4.03 & 3.10 & 0.830 & 8.94 & 4.01\\
2.0 & 2.03 & 1.68 & 0.500 & 4.84 & 2.92 & 3.06 & 2.35 & 0.647 & 6.03 & 3.51\\

\toprule
    $Y_0=0.5$ & \multicolumn{5}{c}{$g_{14}=0.50 ~(0.5\,M_{\odot})$}& \multicolumn{5}{c}{$g_{14}=1.01~(1.0\,M_{\odot})$}\\
    \cmidrule(r){2-6}\cmidrule(r){7-11}
    $Q_{\rm b}(\rm MeV/u)$ & $f(\rm mHz)$ & $f_{\infty}(\rm mHz)$ & $\dot{m}/\dot{m}_{\rm Edd}$ & $L_{36}(\rm max)$ & $L_{36}(\rm min)$
      & $f(\rm mHz)$ & $f_{\infty}(\rm mHz)$ & $\dot{m}/\dot{m}_{\rm Edd}$  & $L_{36}(\rm max)$ & $L_{36}(\rm min)$\\
        \midrule
0 & 1.93 & 1.80 & 0.639 & 6.57 & 3.42 & 4.35 & 3.82 & 1.033 & 10.82 & 5.40 \\
0.15 & 1.87 & 1.75 & 0.612 & 6.39 & 3.24 & 4.19 & 3.68 & 0.994 & 10.21 & 5.33\\
0.5 & 1.70 & 1.59 & 0.550 & 5.58 & 3.01 & 3.86 & 3.39 & 0.891 & 9.36 & 4.76\\
1.0 & 1.46 & 1.36 & 0.456 & 4.69 & 2.46 & 3.19 & 2.80 & 0.745 & 8.49 & 3.71\\
1.5 & 1.12 & 1.05 & 0.360 & 3.54 & 2.03 & 2.70 & 2.37 & 0.605 & 6.35 & 3.25\\
2.0 & - & - & - & - & - & 1.96 & 1.72 & 0.460 & 4.45 & 2.64\\
\toprule
    \multirow{2}{*}{} & \multicolumn{5}{c}{$g_{14}=1.50 ~(1.5\,M_{\odot})$}& \multicolumn{5}{c}{$g_{14}=2.12~(2.0\,M_{\odot})$}\\
    \cmidrule(r){2-6}\cmidrule(r){7-11}
     & $f(\rm mHz)$ & $f_{\infty}(\rm mHz)$ & $\dot{m}/\dot{m}_{\rm Edd}$ & $L_{36}(\rm max)$ & $L_{36}(\rm min)$
      & $f(\rm mHz)$ & $f_{\infty}(\rm mHz)$ & $\dot{m}/\dot{m}_{\rm Edd}$  & $L_{36}(\rm max)$ & $L_{36}(\rm min)$\\
        \midrule
0 & 7.09 & 5.86 & 1.353 & 14.74 & 7.05 & 10.48 & 8.06 & 1.720 & 17.24 & 8.61\\
0.15 & 6.56 & 5.42 & 1.302 & 13.93 & 6.81 & 9.36 & 7.20 & 1.651 & 16.78 & 8.06\\
0.5 & 6.27 & 5.18 & 1.175 & 12.77 & 6.16 & 8.82 & 6.78 & 1.498 & 15.01 & 7.42\\
1.0 & 5.50 & 5.55 & 0.996 & 10.42 & 5.45 & 7.86 & 6.05 & 1.262 & 13.01 & 6.23\\
1.5 & 4.52 & 3.74 & 0.81 & 8.12 & 4.58 & 6.27 & 4.82 & 1.035 & 10.02 & 5.40\\
2.0 & 3.10 & 2.56 & 0.615 & 6.35 & 3.45 & 4.75 & 3.65 & 0.795 & 7.64 & 4.20\\
        \bottomrule
    \end{tabular}
    \label{tab:togashi}
\end{table*}

\begin{table*}
    \centering
    \begin{tabular}{ccccccccccc}
\toprule
    $Y_0=0.6$ & \multicolumn{5}{c}{$g_{14}=0.50 ~(0.5\,M_{\odot})$}& \multicolumn{5}{c}{$g_{14}=1.01~(1.0\,M_{\odot})$}\\
    \cmidrule(r){2-6}\cmidrule(r){7-11}
    $Q_{\rm b}(\rm MeV/u)$ & $f(\rm mHz)$ & $f_{\infty}(\rm mHz)$ & $\dot{m}/\dot{m}_{\rm Edd}$ & $L_{36}(\rm max)$ & $L_{36}(\rm min)$
      & $f(\rm mHz)$ & $f_{\infty}(\rm mHz)$ & $\dot{m}/\dot{m}_{\rm Edd}$  & $L_{36}(\rm max)$ & $L_{36}(\rm min)$\\
        \midrule
0 & 2.73 & 2.55 & 0.770 & 8.25 & 3.96 & 6.80 & 5.96 & 1.245 & 12.80 & 6.67\\
0.15 & 2.69 & 2.51 & 0.74 & 7.81 & 3.89 & 6.15 & 5.39 & 1.196 & 12.50 & 6.33\\
0.5 & 2.43 & 2.27 & 0.665 & 6.99 & 3.50 & 5.79 & 5.08 & 1.081 & 11.28 & 5.75\\
1.0 & 2.21 & 2.07 & 0.555 & 5.80 & 2.94 & 4.82 & 4.23 & 0.915 & 9.49 & 4.85\\
1.5 & 1.81 & 1.69 & 0.440 & 4.65 & 2.33 & 3.70 & 3.25 & 0.740 & 7.32 & 4.16\\
2.0 & - & - & - & - & - & 2.93 & 2.57 & 0.560 & 5.69 & 3.11\\

\toprule
    \multirow{2}{*}{} & \multicolumn{5}{c}{$g_{14}=1.50 ~(1.5\,M_{\odot})$}& \multicolumn{5}{c}{$g_{14}=2.12~(2.0\,M_{\odot})$}\\
    \cmidrule(r){2-6}\cmidrule(r){7-11}
     & $f(\rm mHz)$ & $f_{\infty}(\rm mHz)$ & $\dot{m}/\dot{m}_{\rm Edd}$ & $L_{36}(\rm max)$ & $L_{36}(\rm min)$
      & $f(\rm mHz)$ & $f_{\infty}(\rm mHz)$ & $\dot{m}/\dot{m}_{\rm Edd}$  & $L_{36}(\rm max)$ & $L_{36}(\rm min)$\\
        \midrule
0 & 10.35 & 8.55 & 1.630 & 17.13 & 8.68 & 15.29 & 11.76 & 2.067 & 20.47 & 10.35\\
0.15 & 9.75 & 8.06 & 1.572 & 15.89 & 8.75 & 14.75 & 11.35 & 1.990 & 19.86 & 9.93\\
0.5 & 8.46 & 6.99 & 1.42 & 15.03 & 7.68 & 13.23 & 10.18 & 1.800 & 18.05 & 8.87\\
1.0 & 7.47 & 6.17 & 1.200 & 12.61 & 6.45 & 11.34 & 8.72 & 1.531 & 15.06 & 7.87\\
1.5 & 6.31 & 5.21 & 0.980 & 9.90 & 5.53 & 9.16 & 7.05 & 1.25 & 11.37 & 6.88\\
2.0 & 4.71 & 3.89 & 0.755 & 7.23 & 4.52 & 7.03 & 5.41 & 0.970 & 8.77 & 5.43\\ 

\toprule
    $Y_0=0.7$ & \multicolumn{5}{c}{$g_{14}=0.50 ~(0.5\,M_{\odot})$}& \multicolumn{5}{c}{$g_{14}=1.01~(1.0\,M_{\odot})$}\\
    \cmidrule(r){2-6}\cmidrule(r){7-11}
    $Q_{\rm b}(\rm MeV/u)$ & $f(\rm mHz)$ & $f_{\infty}(\rm mHz)$ & $\dot{m}/\dot{m}_{\rm Edd}$ & $L_{36}(\rm max)$ & $L_{36}(\rm min)$
      & $f(\rm mHz)$ & $f_{\infty}(\rm mHz)$ & $\dot{m}/\dot{m}_{\rm Edd}$  & $L_{36}(\rm max)$ & $L_{36}(\rm min)$\\
        \midrule
0 & 4.02 & 3.76 & 0.920 & 10.41 & 4.44 & 8.82 & 7.74 & 1.486 & 15.66 & 7.76\\
0.15 & 3.70 & 3.46 & 0.885 & 9.17 & 4.68 & 8.50 & 7.46 & 1.428 & 15.02 & 7.54\\
0.5 & 3.61 & 3.37 & 0.800 & 8.12 & 4.37 & 7.65 & 6.71 & 1.295 & 13.17 & 7.01\\
1.0 & 3.03 & 2.83 & 0.671 & 6.49 & 3.83 & 6.94 & 6.09 & 1.095 & 11.09 & 6.01\\
1.5 & 2.36 & 2.01 & 0.535 & 5.60 & 2.88 & 5.75 & 5.04 & 0.89 & 8.94 & 4.94\\
2.0 & - & - & - & - & - & 4.14 & 3.63 & 0.68 & 6.68 & 3.85\\
\toprule
    \multirow{2}{*}{} & \multicolumn{5}{c}{$g_{14}=1.50 ~(1.5\,M_{\odot})$}& \multicolumn{5}{c}{$g_{14}=2.12~(2.0\,M_{\odot})$}\\
    \cmidrule(r){2-6}\cmidrule(r){7-11}
     & $f(\rm mHz)$ & $f_{\infty}(\rm mHz)$ & $\dot{m}/\dot{m}_{\rm Edd}$ & $L_{36}(\rm max)$ & $L_{36}(\rm min)$
      & $f(\rm mHz)$ & $f_{\infty}(\rm mHz)$ & $\dot{m}/\dot{m}_{\rm Edd}$  & $L_{36}(\rm max)$ & $L_{36}(\rm min)$\\
        \midrule
0 & 14.37 & 11.88 & 1.945 & 20.74 & 10.34 & 20.08 & 15.45 & 2.466 & 25.62 & 11.76\\
0.15 & 13.44 & 11.11 & 1.871 & 19.85 & 9.91 & 19.16 & 14.74 & 2.375 & 24.26 & 11.61\\
0.5 & 12.92 & 10.68 & 1.700 & 17.79 & 9.21 & 18.32 & 14.09 & 2.16 & 20.94 & 11.19\\
1.0 & 10.82 & 8.94 & 1.442 & 14.88 & 7.94 & 16.50 & 12.69 & 1.83 & 18.23 & 9.26\\
1.5 & 8.77 & 7.25 & 1.175 & 11.34 & 6.90 & 13.02 & 10.02 & 1.500 & 13.72 & 8.24 \\
2.0 & 6.41 & 5.30 & 0.905 & 9.29 & 5.12 & 10.22 & 7.86 & 1.160 & 11.20 & 6.12\\

\toprule
    $Y_0=0.8$ & \multicolumn{5}{c}{$g_{14}=0.50 ~(0.5\,M_{\odot})$}& \multicolumn{5}{c}{$g_{14}=1.01~(1.0\,M_{\odot})$}\\
    \cmidrule(r){2-6}\cmidrule(r){7-11}
    $Q_{\rm b}(\rm MeV/u)$ & $f(\rm mHz)$ & $f_{\infty}(\rm mHz)$ & $\dot{m}/\dot{m}_{\rm Edd}$ & $L_{36}(\rm max)$ & $L_{36}(\rm min)$
      & $f(\rm mHz)$ & $f_{\infty}(\rm mHz)$ & $\dot{m}/\dot{m}_{\rm Edd}$  & $L_{36}(\rm max)$ & $L_{36}(\rm min)$\\
        \midrule
0 & 5.52 & 5.16 & 1.095 & 11.25 & 5.80 & 12.82 & 11.25 & 1.768 & 18.10 & 9.44\\
0.15 & 4.98 & 4.65 & 1.051 & 11.91 & 5.08 & 11.19 & 9.82 & 1.700 & 17.67 & 9.02\\
0.5 & 4.55 & 4.25 & 0.950 & 10.15 & 4.98 & 10.62 & 9.32 & 1.540 & 16.21 & 8.06\\
1.0 & 4.40 & 4.11 & 0.802 & 8.27 & 4.34 & 9.42 & 8.26 & 1.305 & 12.76 & 7.45\\
1.5 & 3.26 & 2.86 & 0.650 & 6.41 & 3.68 & 7.75 & 6.80 & 1.065 & 10.26 & 6.16\\
2.0 & - & - & - & - & - & 5.81 & 5.10 & 0.82 & 8.00 & 4.65\\
\toprule
    \multirow{2}{*}{} & \multicolumn{5}{c}{$g_{14}=1.50 ~(1.5\,M_{\odot})$}& \multicolumn{5}{c}{$g_{14}=2.12~(2.0\,M_{\odot})$}\\
    \cmidrule(r){2-6}\cmidrule(r){7-11}
     & $f(\rm mHz)$ & $f_{\infty}(\rm mHz)$ & $\dot{m}/\dot{m}_{\rm Edd}$ & $L_{36}(\rm max)$ & $L_{36}(\rm min)$
      & $f(\rm mHz)$ & $f_{\infty}(\rm mHz)$ & $\dot{m}/\dot{m}_{\rm Edd}$  & $L_{36}(\rm max)$ & $L_{36}(\rm min)$\\
        \midrule
0 & 18.73 & 15.48 & 2.315 & 26.90 & 11.09 & 28.25 & 21.73 & 2.945 & 28.44 & 15.08\\
0.15 & 18.52 & 15.31 & 2.232 & 23.77 & 11.66 & 27.32 & 21.02 & 2.835 & 27.85 & 14.37\\
0.5 & 17.54 & 14.50 & 2.020 & 22.00 & 10.52 & 25.64 & 19.72 & 2.570 & 25.93 & 12.72\\
1.0 & 15.43 & 12.75 & 1.720 & 18.24 & 9.28 & 21.93 & 16.87 & 2.178 &21.28 & 11.29\\
1.5 & 12.53 & 10.36 & 1.410 & 14.58 & 7.80 & 19.61 & 15.08 & 1.805 & 17.06 & 9.50\\
2.0 & 9.21 & 7.61 & 1.090 & 11.01 & 6.19 & 13.66 & 10.51 & 1.400 & 12.92 & 7.55\\

        \bottomrule
    \end{tabular}
    
\end{table*}

\begin{table*}
    \centering
    \begin{tabular}{ccccccccccc}
\toprule
    $Y_0=0.9$ & \multicolumn{5}{c}{$g_{14}=0.50 ~(0.5\,M_{\odot})$}& \multicolumn{5}{c}{$g_{14}=1.01~(1.0\,M_{\odot})$}\\
    \cmidrule(r){2-6}\cmidrule(r){7-11}
    $Q_{\rm b}(\rm MeV/u)$ & $f(\rm mHz)$ & $f_{\infty}(\rm mHz)$ & $\dot{m}/\dot{m}_{\rm Edd}$ & $L_{36}(\rm max)$ & $L_{36}(\rm min)$
      & $f(\rm mHz)$ & $f_{\infty}(\rm mHz)$ & $\dot{m}/\dot{m}_{\rm Edd}$  & $L_{36}(\rm max)$ & $L_{36}(\rm min)$\\
        \midrule
0 & 7.58 & 7.08 & 1.300 & 13.57 & 6.90 & 16.18 & 14.19 & 2.105 & 21.70 & 11.33\\
0.15 & 6.69 & 6.25 & 1.250 & 12.99 & 6.62 & 16.13 & 14.15 & 2.025 & 21.33 & 10.60\\
0.5 & 6.41 & 5.99 & 1.130 & 11.66 & 6.04 & 14.01 & 12.29 & 1.835 & 19.60 & 9.49\\
1.0 & 5.89 & 5.50 & 0.960 & 9.82 & 5.25 & 13.44 & 11.79 & 1.565 & 16.32 & 8.40\\
1.5 & 4.47 & 4.18 & 0.780 & 7.87 & 4.28 & 10.96 & 9.61 & 1.280 & 13.19 & 7.00\\
2.0 & - & - & - & - & - & 8.09 & 7.10 & 0.99 & 9.57 & 5.65\\

\toprule
    \multirow{2}{*}{} & \multicolumn{5}{c}{$g_{14}=1.50 ~(1.5\,M_{\odot})$}& \multicolumn{5}{c}{$g_{14}=2.12~(2.0\,M_{\odot})$}\\
    \cmidrule(r){2-6}\cmidrule(r){7-11}
     & $f(\rm mHz)$ & $f_{\infty}(\rm mHz)$ & $\dot{m}/\dot{m}_{\rm Edd}$ & $L_{36}(\rm max)$ & $L_{36}(\rm min)$
      & $f(\rm mHz)$ & $f_{\infty}(\rm mHz)$ & $\dot{m}/\dot{m}_{\rm Edd}$  & $L_{36}(\rm max)$ & $L_{36}(\rm min)$\\
        \midrule
0 & 26.04 & 21.52 & 2.765 & 31.71 & 13.44 & 38.76 & 29.82 & 3.519 & 36.43 & 16.87\\
0.15 & 25.64 & 21.19 & 2.664 & 27.65 & 14.41 & 37.04 & 28.49 & 3.391 & 33.10 & 17.25\\
0.5 & 23.81 & 19.68 & 2.420 & 25.10 & 13.18 & 36.23 & 27.87 & 3.090 & 27.34 & 17.22\\
1.0 & 20.33 & 16.80 & 2.066 & 19.95 & 12.11 & 31.45 & 24.19 & 2.630 & 25.96 & 13.40\\
1.5 & 16.84 & 13.92 & 1.69 & 17.83 & 9.21 & 24.88 & 19.14 & 2.165 & 19.72 & 11.96\\
2.0 & 12.08 & 9.98 & 1.325 & 13.18 & 7.69 & 18.52 & 14.25 & 1.699 & 15.71 & 9.23\\

        \bottomrule
    \end{tabular}
    \label{tab:TogashiA}
\end{table*}


\begin{table*}
    \centering
    \caption{The same as table~\ref{tab:togashi} but the surface gravities are inferred from MIT EoS.
    }
    \begin{tabular}{ccccccccccc}
    \toprule
    $Y_0=0.3$ & \multicolumn{5}{c}{$g_{14}=1.15 ~(0.5\,M_{\odot})$}& \multicolumn{5}{c}{$g_{14}=1.51~(1.0\,M_{\odot})$}\\
    \cmidrule(r){2-6}\cmidrule(r){7-11}
    $Q_{\rm b}(\rm MeV/u)$ & $f(\rm mHz)$ & $f_{\infty}(\rm mHz)$ & $\dot{m}/\dot{m}_{\rm Edd}$ & $L_{36}(\rm max)$ & $L_{36}(\rm min)$
      & $f(\rm mHz)$ & $f_{\infty}(\rm mHz)$ & $\dot{m}/\dot{m}_{\rm Edd}$  & $L_{36}(\rm max)$ & $L_{36}(\rm min)$\\
        \midrule
0 & 2.20 & 2.00 & 0.732 & 3.35 & 1.71 & 2.80 & 2.39 & 0.889 & 5.99 & 3.24\\
0.15 & 1.96 & 1.78 & 0.701 & 3.29 & 1.60 & 2.61 & 2.23 & 0.851 & 5.95 & 2.99\\
0.5 & 1.73 & 1.57 & 0.635 & 2.82 & 1.54 & 2.36 & 2.02 & 0.765 & 5.54 & 2.62\\
1.0 & 1.49 & 1.35 & 0.530 & 2.32 & 1.30 & 2.20 & 1.88 & 0.645 & 4.45 & 2.30\\
1.5 & 1.17 & 1.06 & 0.415 & 1.98 & 0.96 & 1.70 & 1.45 & 0.520 & 3.42 & 1.96\\
2.0 & 0.84 & 0.76 & 0.305 & 1.31 & 0.76 & 1.21 & 1.03 & 0.39 & 2.58 & 1.50\\

\toprule
    \multirow{2}{*}{} & \multicolumn{5}{c}{$g_{14}=1.88 ~(1.5\,M_{\odot})$}& \multicolumn{5}{c}{$g_{14}=2.69~(2.0\,M_{\odot})$}\\
    \cmidrule(r){2-6}\cmidrule(r){7-11}
     & $f(\rm mHz)$ & $f_{\infty}(\rm mHz)$ & $\dot{m}/\dot{m}_{\rm Edd}$ & $L_{36}(\rm max)$ & $L_{36}(\rm min)$
      & $f(\rm mHz)$ & $f_{\infty}(\rm mHz)$ & $\dot{m}/\dot{m}_{\rm Edd}$  & $L_{36}(\rm max)$ & $L_{36}(\rm min)$\\
        \midrule
0 & 3.59 & 2.90 & 1.035 & 8.58 & 4.44 & 5.63 & 4.20 & 1.329 & 10.47 & 5.23\\
0.15 & 3.23 & 2.60 & 0.995 & 8.28 & 4.31 & 5.43 & 4.05 & 1.278 & 10.09 & 5.06\\
0.5 & 2.94 & 2.37 & 0.900 & 7.42 & 3.95 & 4.52 & 3.37 & 1.16 & 8.91 & 4.69\\
1.0 & 2.52 & 2.03 & 0.755 & 6.39 & 3.24 & 3.71 & 2.77 & 0.980 & 7.56 & 3.98\\
1.5 & 2.13 & 1.72 & 0.61 & 5.05 & 2.67 & 3.09 & 2.31 & 0.797 & 5.90 & 3.33\\
2.0 & 1.51 & 1.22 & 0.460 & 3.71 & 2.08 & 2.36 & 1.76 & 0.605 & 4.00 & 2.84\\

\toprule
    $Y_0=0.4$ & \multicolumn{5}{c}{$g_{14}=1.15 ~(0.5\,M_{\odot})$}& \multicolumn{5}{c}{$g_{14}=1.51~(1.0\,M_{\odot})$}\\
    \cmidrule(r){2-6}\cmidrule(r){7-11}
    $Q_{\rm b}(\rm MeV/u)$ & $f(\rm mHz)$ & $f_{\infty}(\rm mHz)$ & $\dot{m}/\dot{m}_{\rm Edd}$ & $L_{36}(\rm max)$ & $L_{36}(\rm min)$
      & $f(\rm mHz)$ & $f_{\infty}(\rm mHz)$ & $\dot{m}/\dot{m}_{\rm Edd}$  & $L_{36}(\rm max)$ & $L_{36}(\rm min)$\\
        \midrule
0 & 3.46 & 3.15 & 0.923 & 4.24 & 2.15 & 4.49 & 3.84 & 1.112 & 8.06 & 3.77\\
0.15 & 3.31 & 3.01 & 0.888 & 4.02 & 2.09 & 4.23 & 3.62 & 1.070 & 7.56 & 3.77\\
0.5 & 2.83 & 2.57 & 0.798 & 3.73 & 1.84 & 3.82 & 3.26 & 0.960 & 7.73 & 2.99\\
1.0 & 2.59 & 2.35 & 0.673 & 2.99 & 1.61 & 3.35 & 2.86 & 0.810 & 6.14 & 2.70\\
1.5 & 2.01 & 1.83 & 0.540 & 2.43 & 1.31 & 2.65 & 2.26 & 0.66 & 4.50 & 2.42\\
2.0 & 1.53 & 1.39 & 0.405 & 1.77 & 1.01 & 2.03 & 1.74 & 0.5 & 3.36 & 1.88\\
\toprule
    \multirow{2}{*}{} & \multicolumn{5}{c}{$g_{14}=1.88 ~(1.5\,M_{\odot})$}& \multicolumn{5}{c}{$g_{14}=2.69~(2.0\,M_{\odot})$}\\
    \cmidrule(r){2-6}\cmidrule(r){7-11}
     & $f(\rm mHz)$ & $f_{\infty}(\rm mHz)$ & $\dot{m}/\dot{m}_{\rm Edd}$ & $L_{36}(\rm max)$ & $L_{36}(\rm min)$
      & $f(\rm mHz)$ & $f_{\infty}(\rm mHz)$ & $\dot{m}/\dot{m}_{\rm Edd}$  & $L_{36}(\rm max)$ & $L_{36}(\rm min)$\\
        \midrule
0 & 5.93 & 4.78 & 1.295 & 11.03 & 5.42 & 8.77 & 6.54 & 1.660 & 13.27 & 6.45\\
0.15 & 5.69 & 4.59 & 1.247 & 10.29 & 5.40 & 8.38 & 6.25 & 1.598 & 12.65 & 6.25\\
0.5 & 5.18 & 4.18 & 1.13 & 9.36 & 4.93 & 7.41 & 5.53 & 1.445 & 11.80 & 5.62\\
1.0 & 4.41 & 3.56 & 0.955 & 7.81 & 4.18 & 6.41 & 4.78 & 1.225 & 9.70 & 4.89\\
1.5 & 3.60 & 2.90 & 0.770 & 5.74 & 3.67 & 5.38 & 4.01 & 1.000 & 7.28 & 4.31\\
2.0 & 2.65 & 2.14 & 0.590 & 4.74 & 2.68 & 3.98 & 2.97 & 0.766 & 5.87 & 3.19\\

\toprule
    $Y_0=0.5$ & \multicolumn{5}{c}{$g_{14}=1.15 ~(0.5\,M_{\odot})$}& \multicolumn{5}{c}{$g_{14}=1.51~(1.0\,M_{\odot})$}\\
    \cmidrule(r){2-6}\cmidrule(r){7-11}
    $Q_{\rm b}(\rm MeV/u)$ & $f(\rm mHz)$ & $f_{\infty}(\rm mHz)$ & $\dot{m}/\dot{m}_{\rm Edd}$ & $L_{36}(\rm max)$ & $L_{36}(\rm min)$
      & $f(\rm mHz)$ & $f_{\infty}(\rm mHz)$ & $\dot{m}/\dot{m}_{\rm Edd}$  & $L_{36}(\rm max)$ & $L_{36}(\rm min)$\\
        \midrule
0 & 4.98 & 4.53 & 1.131 & 5.11 & 2.66 & 7.34 & 6.27 & 1.361 & 9.40 & 4.84\\
0.15 & 4.99 & 4.54 & 1.086 & 5.02 & 2.54 & 7.22 & 6.17 & 1.310 & 9.12 & 4.62\\
0.5 & 4.33 & 3.94 & 0.980 & 4.57 & 2.28 & 6.17 & 5.27 & 1.185 & 8.19 & 4.26\\
1.0 & 3.85 & 3.5 & 0.825 & 3.76 & 1.94 & 5.18 & 4.43 & 1.000 & 7.12 & 3.56\\
1.5 & 3.26 & 2.96 & 0.670 & 2.96 & 1.66 & 4.36 & 3.73 & 0.815 & 5.43 & 3.03\\
2.0 & 2.24 & 2.04 & 0.510 & 2.14 & 1.30 & 3.19 & 2.73 & 0.620 & 4.13 & 2.35\\
\toprule
    \multirow{2}{*}{} & \multicolumn{5}{c}{$g_{14}=1.88 ~(1.5\,M_{\odot})$}& \multicolumn{5}{c}{$g_{14}=2.69~(2.0\,M_{\odot})$}\\
    \cmidrule(r){2-6}\cmidrule(r){7-11}
     & $f(\rm mHz)$ & $f_{\infty}(\rm mHz)$ & $\dot{m}/\dot{m}_{\rm Edd}$ & $L_{36}(\rm max)$ & $L_{36}(\rm min)$
      & $f(\rm mHz)$ & $f_{\infty}(\rm mHz)$ & $\dot{m}/\dot{m}_{\rm Edd}$  & $L_{36}(\rm max)$ & $L_{36}(\rm min)$\\
        \midrule
0 & 8.91 & 7.19 & 1.585 & 13.04 & 6.87 & 12.92 & 9.64 & 2.025 & 16.24 & 7.83\\
0.15 & 8.77 & 7.07& 1.523 & 12.59 & 6.58 & 12.17 & 9.08 & 1.945 & 16.01 & 7.36\\
0.5 & 7.82 & 6.31 & 1.370 & 11.26 & 5.98 & 11.82 & 8.82 & 1.765 & 13.09 & 7.39\\
1.0 & 6.51 & 5.25 & 1.150 & 10.77 & 4.52 & 10.04 & 7.49 & 1.501 & 11.68 & 6.10\\
1.5 & 5.65 & 4.56 & 0.950 & 7.83 & 4.22 & 8.21 & 6.13 & 1.230 & 9.28 & 5.15\\
2.0 & 4.11 & 3.31 & 0.730 & 5.68 & 3.35 & 6.20 & 4.63 & 0.950 & 6.90 & 4.08\\
        \bottomrule
    \end{tabular}
\end{table*}

\begin{table*}
    \centering
    \begin{tabular}{ccccccccccc}
    \toprule
    $Y_0=0.6$ & \multicolumn{5}{c}{$g_{14}=1.15 ~(0.5\,M_{\odot})$}& \multicolumn{5}{c}{$g_{14}=1.51~(1.0\,M_{\odot})$}\\
    \cmidrule(r){2-6}\cmidrule(r){7-11}
    $Q_{\rm b}(\rm MeV/u)$ & $f(\rm mHz)$ & $f_{\infty}(\rm mHz)$ & $\dot{m}/\dot{m}_{\rm Edd}$ & $L_{36}(\rm max)$ & $L_{36}(\rm min)$
      & $f(\rm mHz)$ & $f_{\infty}(\rm mHz)$ & $\dot{m}/\dot{m}_{\rm Edd}$  & $L_{36}(\rm max)$ & $L_{36}(\rm min)$\\
        \midrule
0 & 7.03 & 6.39 & 1.360 & 6.17 & 3.20 & 9.47 & 8.09 & 1.635 & 11.69 & 5.61\\
0.15 & 7.41 & 6.74 & 1.308 & 5.96 & 3.03 & 9.36 & 8.00 & 1.575 & 10.99 & 5.53\\
0.5 & 6.56 & 5.96 & 1.18 & 4.92 & 3.01 & 8.91 & 7.62 & 1.420 & 9.56 & 5.17\\
1.0 & 5.63 & 5.12 & 1.000 & 4.50 & 2.40 & 7.22 & 6.17 & 1.210 & 8.16 & 4.44\\
1.5 & 4.47 & 4.06 & 0.810 & 3.67 & 1.93 & 6.29 & 5.38 & 0.980 & 6.65 & 3.60\\
2.0 & 3.39 & 3.08 & 0.620 & 2.63 & 1.56 & 4.68 & 4.00 & 0.760 & 4.95 & 2.93\\

\toprule
    \multirow{2}{*}{} & \multicolumn{5}{c}{$g_{14}=1.88 ~(1.5\,M_{\odot})$}& \multicolumn{5}{c}{$g_{14}=2.69~(2.0\,M_{\odot})$}\\
    \cmidrule(r){2-6}\cmidrule(r){7-11}
     & $f(\rm mHz)$ & $f_{\infty}(\rm mHz)$ & $\dot{m}/\dot{m}_{\rm Edd}$ & $L_{36}(\rm max)$ & $L_{36}(\rm min)$
      & $f(\rm mHz)$ & $f_{\infty}(\rm mHz)$ & $\dot{m}/\dot{m}_{\rm Edd}$  & $L_{36}(\rm max)$ & $L_{36}(\rm min)$\\
        \midrule
0 & 12.72 & 10.26 & 1.900 & 16.62 & 7.72 & 19.38 & 14.46 & 2.439 & 18.89 & 9.81\\
0.15 & 12.17 & 9.81 & 1.827 & 15.91 & 7.46 & 18.73 & 13.98 & 2.351 & 17.85 & 9.70\\
0.5 & 11.90 & 9.60 & 1.660 & 13.77 & 7.22 & 17.54 & 13.09 & 2.130 & 16.40 & 8.59\\
1.0 & 10.16 & 8.19 & 1.405 & 11.75 & 6.04 & 14.37 & 10.72 & 1.811 & 14.06 & 7.33\\
1.5 & 7.75 & 6.25 & 1.125 & 10.58 & 4.50 & 11.99 & 8.95 & 1.48 & 10.62 & 6.45\\
2.0 & 6.13 & 4.94 & 0.89 & 6.94 & 4.11 & 9.16 & 6.84 & 1.15 & 8.20 & 5.10\\

\toprule
    $Y_0=0.7$ & \multicolumn{5}{c}{$g_{14}=1.15 ~(0.5\,M_{\odot})$}& \multicolumn{5}{c}{$g_{14}=1.51~(1.0\,M_{\odot})$}\\
    \cmidrule(r){2-6}\cmidrule(r){7-11}
    $Q_{\rm b}(\rm MeV/u)$ & $f(\rm mHz)$ & $f_{\infty}(\rm mHz)$ & $\dot{m}/\dot{m}_{\rm Edd}$ & $L_{36}(\rm max)$ & $L_{36}(\rm min)$
      & $f(\rm mHz)$ & $f_{\infty}(\rm mHz)$ & $\dot{m}/\dot{m}_{\rm Edd}$  & $L_{36}(\rm max)$ & $L_{36}(\rm min)$\\
        \midrule
0 & 9.92 & 9.02 & 1.621 & 7.63 & 3.68 & 14.62 & 12.50 & 1.955 & 13.38 & 6.98\\
0.15 & 9.98 & 9.07 & 1.561 & 7.11 & 3.63 & 13.44 & 11.49 & 1.880 & 13.10 & 6.63\\
0.5 & 9.11 & 8.28 & 1.41 & 6.72 & 3.18 & 12.25 & 10.47 & 1.700 & 12.59 & 5.76\\
1.0 & 8.25 & 7.50 & 1.195 & 5.41 & 2.83 & 9.69 & 8.28 & 1.420 & 11.87 & 4.33\\
1.5 & 6.06 & 5.51 & 0.975 & 4.35 & 2.37 & 8.17 & 6.98 & 1.160 & 9.02 & 3.87\\
2.0 & 4.85 & 4.41 & 0.750 & 3.16 & 1.90 & 6.61 & 5.65 & 0.91 & 5.97 & 3.43\\
\toprule
    \multirow{2}{*}{} & \multicolumn{5}{c}{$g_{14}=1.88 ~(1.5\,M_{\odot})$}& \multicolumn{5}{c}{$g_{14}=2.69~(2.0\,M_{\odot})$}\\
    \cmidrule(r){2-6}\cmidrule(r){7-11}
     & $f(\rm mHz)$ & $f_{\infty}(\rm mHz)$ & $\dot{m}/\dot{m}_{\rm Edd}$ & $L_{36}(\rm max)$ & $L_{36}(\rm min)$
      & $f(\rm mHz)$ & $f_{\infty}(\rm mHz)$ & $\dot{m}/\dot{m}_{\rm Edd}$  & $L_{36}(\rm max)$ & $L_{36}(\rm min)$\\
        \midrule
0 & 18.12 & 14.61 & 2.270 & 17.95 & 10.24 & 28.74 & 21.45 & 2.915 & 22.33 & 11.71\\
0.15 & 16.84 & 13.58 & 2.185 & 18.92 & 9.00 & 26.04 & 19.43 & 2.800 & 21.87 & 11.11\\
0.5 & 16.34 & 13.18 & 1.980 & 16.70 & 8.39 & 23.15 & 17.28 & 2.545 & 19.81 & 10.21\\
1.0 & 13.02 & 10.50 & 1.655 & 16.25 & 6.20 & 21.37 & 11.95 & 2.170 & 16.54 & 8.82\\
1.5 & 11.34 & 9.15 & 1.365 & 12.11 & 5.70 & 17.01 & 12.69 & 1.775 & 12.90 & 7.68\\
2.0 & 8.55 & 6.90 & 1.055 & 9.08 & 4.57 & 12.82 & 9.57 & 1.39 & 10.12 & 5.98\\

\toprule
    $Y_0=0.8$ & \multicolumn{5}{c}{$g_{14}=1.15 ~(0.5\,M_{\odot})$}& \multicolumn{5}{c}{$g_{14}=1.51~(1.0\,M_{\odot})$}\\
    \cmidrule(r){2-6}\cmidrule(r){7-11}
    $Q_{\rm b}(\rm MeV/u)$ & $f(\rm mHz)$ & $f_{\infty}(\rm mHz)$ & $\dot{m}/\dot{m}_{\rm Edd}$ & $L_{36}(\rm max)$ & $L_{36}(\rm min)$
      & $f(\rm mHz)$ & $f_{\infty}(\rm mHz)$ & $\dot{m}/\dot{m}_{\rm Edd}$  & $L_{36}(\rm max)$ & $L_{36}(\rm min)$\\
        \midrule
0 & 14.25 & 12.95 & 1.930 & 9.31 & 4.29 & 19.61 & 16.76 & 2.33 & 15.89 & 8.35\\
0.15 & 14.01 & 12.74 & 1.860 & 8.24 & 4.46 & 18.12 & 15.49 & 2.239 & 15.87 & 7.79\\
0.5 & 13.23 & 12.03 & 1.685 & 7.72 & 3.92 & 17.73 & 15.15 & 2.03 & 13.17 & 7.66 \\
1.0 & 11.26 & 10.24 & 1.430 & 6.47 & 3.40 & 14.37 & 12.28 & 1.710 & 13.17 & 5.62 \\
1.5 & 8.96 & 8.15 & 1.165 & 5.30 & 2.77 & 12.17 & 10.40 & 1.420 & 9.65 & 5.19 \\
2.0 & 6.75 & 6.14 & 0.900 & 3.88 & 2.23 & 8.68 & 7.42 & 1.100 & 7.20 & 4.19 \\
\toprule
    \multirow{2}{*}{} & \multicolumn{5}{c}{$g_{14}=1.88 ~(1.5\,M_{\odot})$}& \multicolumn{5}{c}{$g_{14}=2.69~(2.0\,M_{\odot})$}\\
    \cmidrule(r){2-6}\cmidrule(r){7-11}
     & $f(\rm mHz)$ & $f_{\infty}(\rm mHz)$ & $\dot{m}/\dot{m}_{\rm Edd}$ & $L_{36}(\rm max)$ & $L_{36}(\rm min)$
      & $f(\rm mHz)$ & $f_{\infty}(\rm mHz)$ & $\dot{m}/\dot{m}_{\rm Edd}$  & $L_{36}(\rm max)$ & $L_{36}(\rm min)$\\
        \midrule
0 & 24.88 & 20.06 & 2.705 & 24.65 & 10.63 & 37.88 & 28.27 & 3.475 & 27.21 & 13.86\\
0.15 & 23.47 & 18.93 & 2.608 & 21.89 & 11.23 & 37.88 & 28.27 & 3.345 & 25.75 & 13.43\\
0.5 & 22.22 & 17.92 & 2.360 & 18.83 & 10.56 & 32.68 & 24.39 & 3.040 & 23.96 & 12.14\\
1.0 & 18.94 & 15.27 & 2.000 & 16.65 & 8.70 & 29.76 & 22.21 & 2.59 & 20.14 & 10.43\\
1.5 & 14.49 & 11.69 & 1.62 & 15.18 & 6.46 & 23.15 & 17.28 & 2.11 & 17.24 & 8.22\\
2.0 & 11.99 & 9.67 & 1.290 & 9.84 & 6.03 & 18.52 & 13.82 & 1.670 & 11.74 & 7.49\\

        \bottomrule
    \end{tabular}
\end{table*}

\begin{table*}
    \centering
    \begin{tabular}{ccccccccccc}
    \toprule
    $Y_0=0.9$ & \multicolumn{5}{c}{$g_{14}=1.15 ~(0.5\,M_{\odot})$}& \multicolumn{5}{c}{$g_{14}=1.51~(1.0\,M_{\odot})$}\\
    \cmidrule(r){2-6}\cmidrule(r){7-11}
    $Q_{\rm b}(\rm MeV/u)$ & $f(\rm mHz)$ & $f_{\infty}(\rm mHz)$ & $\dot{m}/\dot{m}_{\rm Edd}$ & $L_{36}(\rm max)$ & $L_{36}(\rm min)$
      & $f(\rm mHz)$ & $f_{\infty}(\rm mHz)$ & $\dot{m}/\dot{m}_{\rm Edd}$  & $L_{36}(\rm max)$ & $L_{36}(\rm min)$\\
        \midrule
0 & 19.61 & 17.83 & 2.301 & 10.51 & 5.36 & 26.04 & 22.26 & 2.778 & 21.96 & 8.63\\
0.15 & 18.94 & 17.22 & 2.215 & 9.97 & 5.20 & 25.25 & 21.58 & 2.675 & 18.74 & 9.36\\
0.5 & 17.36 & 15.78 & 2.010 & 9.32 & 4.64 & 23.15 & 19.79 & 2.430 & 16.98 & 8.57\\
1.0 & 15.29 & 13.90 & 1.710 & 7.89 & 4.03 & 20.33 & 17.38 & 2.060 & 13.97 & 7.60\\
1.5 & 12.44 & 11.31 & 1.400 & 6.44 & 3.33 & 16.67 & 14.25 & 1.680 & 12.11 & 5.92\\
2.0 & 9.36 & 8.51 & 1.090 & 4.61 & 2.75 & 13.12 & 11.21 & 1.330 & 8.38 & 5.27\\

\toprule
    \multirow{2}{*}{} & \multicolumn{5}{c}{$g_{14}=1.88 ~(1.5\,M_{\odot})$}& \multicolumn{5}{c}{$g_{14}=2.69~(2.0\,M_{\odot})$}\\
    \cmidrule(r){2-6}\cmidrule(r){7-11}
     & $f(\rm mHz)$ & $f_{\infty}(\rm mHz)$ & $\dot{m}/\dot{m}_{\rm Edd}$ & $L_{36}(\rm max)$ & $L_{36}(\rm min)$
      & $f(\rm mHz)$ & $f_{\infty}(\rm mHz)$ & $\dot{m}/\dot{m}_{\rm Edd}$  & $L_{36}(\rm max)$ & $L_{36}(\rm min)$\\
        \midrule
0 & 32.05 & 25.85 & 3.235 & 28.64 & 12.93 & 49.02 & 36.58 & 4.155 & 34.11 & 15.72\\
0.15 & 32.05 & 25.85 & 3.119 & 26.07 & 13.19 & 47.62 & 35.54 & 4.01 & 30.52 & 16.41\\
0.5 & 30.86 & 24.89 & 2.837 & 23.76 & 12.18 & 45.05 & 33.62 & 3.652 & 28.58 & 14.38\\
1.0 & 26.46 & 21.34 & 2.410 & 19.29 & 10.47 & 39.68 & 29.61 & 3.100 & 23.16 & 13.06\\
1.5 & 20.58 & 15.36 & 1.920 & 20.03 & 7.00 & 30.86 & 23.03 & 2.500 & 22.15 & 9.08\\
2.0 & 15.87 & 12.80 & 1.485 & 15.40 & 5.78 & 25.25 & 18.84 & 2.010 & 14.00 & 9.02\\

        \bottomrule
    \end{tabular}
    \label{tab:MITA}
\end{table*}

\begin{table*}
    \centering
    \caption{The same as table~\ref{tab:togashi} but the surface gravities are inferred from LX3630 EoS. As the surface gravity of $1.5M_{\odot}$ is the same as that from MIT EoS, we leave it out here.  
    }
    \begin{tabular}{ccccccccccc}
    \toprule
    $Y_0=0.3$ & \multicolumn{5}{c}{$g_{14}=1.22 ~(0.5\,M_{\odot})$}& \multicolumn{5}{c}{$g_{14}=1.59~(1.0\,M_{\odot})$}\\
    \cmidrule(r){2-6}\cmidrule(r){7-11}
    $Q_{\rm b}(\rm MeV/u)$ & $f(\rm mHz)$ & $f_{\infty}(\rm mHz)$ & $\dot{m}/\dot{m}_{\rm Edd}$ & $L_{36}(\rm max)$ & $L_{36}(\rm min)$
      & $f(\rm mHz)$ & $f_{\infty}(\rm mHz)$ & $\dot{m}/\dot{m}_{\rm Edd}$  & $L_{36}(\rm max)$ & $L_{36}(\rm min)$\\
        \midrule
0 & 2.32 & 2.09 & 0.765 & 3.19 & 1.72 & 2.79 & 2.38 & 0.915 & 7.29 & 2.56\\
0.15 & 2.12 & 1.91 & 0.730 & 3.21 & 1.58 & 2.71 & 2.32 & 0.885 & 5.86 & 3.01\\
0.5 & 1.96 & 1.77 & 0.660 & 2.84 & 1.47 & 2.47 & 2.11 & 0.800 & 5.12 & 2.81\\
1.0 & 1.59 & 1.43 & 0.551 & 2.38 & 1.21 & 2.16 & 1.85 & 0.670 & 4.40 & 2.30\\
1.5 & 1.30 & 1.17 & 0.440 & 1.83 & 1.02 & 1.79 & 1.53 & 0.540 & 3.41 & 1.92\\
2.0 & 0.85 & 0.77 & 0.32 & 1.37 & 0.73 & 1.28 & 1.09 & 0.405 & 2.50 & 1.49\\

\toprule
    \multirow{2}{*}{} & \multicolumn{5}{c}{$g_{14}=2.16 ~(2.0\,M_{\odot})$}& \multicolumn{5}{c}{$g_{14}=2.86~(3.0\,M_{\odot})$}\\
    \cmidrule(r){2-6}\cmidrule(r){7-11}
     & $f(\rm mHz)$ & $f_{\infty}(\rm mHz)$ & $\dot{m}/\dot{m}_{\rm Edd}$ & $L_{36}(\rm max)$ & $L_{36}(\rm min)$
      & $f(\rm mHz)$ & $f_{\infty}(\rm mHz)$ & $\dot{m}/\dot{m}_{\rm Edd}$  & $L_{36}(\rm max)$ & $L_{36}(\rm min)$\\
        \midrule
0 & 4.07 & 3.13 & 1.140 & 10.97 & 5.70 & 5.93 & 4.12 & 1.390 & 15.00 & 7.92\\
0.15 & 3.76 & 2.89 & 1.090 & 11.91 & 4.88 & 5.36 & 3.72 & 1.330 & 13.78 & 7.93\\
0.5 & 3.66 & 2.82 & 0.987 & 10.05 & 4.77 & 4.76 & 3.31 & 1.210 & 13.13 & 6.92\\
1.0 & 3.10 & 2.38 & 0.838 & 7.53 & 5.52 & 4.59 & 3.19 & 1.025 & 10.99 & 5.90\\
1.5 & 2.57 & 1.98 & 0.680 & 6.24 & 3.59 & 3.45 & 2.40 & 0.830 & 8.49 & 5.06\\
2.0 & 1.82 & 1.40 & 0.505 & 5.05 & 2.55 & 2.60 & 1.81 & 0.630 & 6.85 & 3.69\\

\toprule
    $Y_0=0.4$ & \multicolumn{5}{c}{$g_{14}=1.22 ~(0.5\,M_{\odot})$}& \multicolumn{5}{c}{$g_{14}=1.59~(1.0\,M_{\odot})$}\\
    \cmidrule(r){2-6}\cmidrule(r){7-11}
    $Q_{\rm b}(\rm MeV/u)$ & $f(\rm mHz)$ & $f_{\infty}(\rm mHz)$ & $\dot{m}/\dot{m}_{\rm Edd}$ & $L_{36}(\rm max)$ & $L_{36}(\rm min)$
      & $f(\rm mHz)$ & $f_{\infty}(\rm mHz)$ & $\dot{m}/\dot{m}_{\rm Edd}$  & $L_{36}(\rm max)$ & $L_{36}(\rm min)$\\
        \midrule
0 & 3.65 & 3.29 & 0.960 & 4.25 & 2.04 & 4.89 & 4.18 & 1.155 & 7.60 & 3.91\\
0.15 & 3.20 & 2.88 & 0.920 & 4.27 & 1.90 & 4.72 & 4.03 & 1.11 & 7.46 & 3.74\\
0.5 & 3.07 & 2.77 & 0.830 & 3.76 & 1.74 & 4.42 & 3.78 & 1.000 & 6.92 & 3.25\\
1.0 & 2.66 & 2.40 & 0.695 & 3.16 & 1.46 & 3.62 & 3.09 & 0.845 & 5.64 & 2.87\\
1.5 & 2.10 & 1.89 & 0.545 & 2.81 & 1.07 & 2.87 & 2.45 & 0.680 & 4.56 & 2.32\\
2.0 & 1.54 & 1.39 & 0.420 & 1.81 & 0.95 & 2.16 & 1.85 & 0.520 & 2.91 & 2.07\\
\toprule
    \multirow{2}{*}{} & \multicolumn{5}{c}{$g_{14}=2.16 ~(2.0\,M_{\odot})$}& \multicolumn{5}{c}{$g_{14}=2.86~(3.0\,M_{\odot})$}\\
    \cmidrule(r){2-6}\cmidrule(r){7-11}
     & $f(\rm mHz)$ & $f_{\infty}(\rm mHz)$ & $\dot{m}/\dot{m}_{\rm Edd}$ & $L_{36}(\rm max)$ & $L_{36}(\rm min)$
      & $f(\rm mHz)$ & $f_{\infty}(\rm mHz)$ & $\dot{m}/\dot{m}_{\rm Edd}$  & $L_{36}(\rm max)$ & $L_{36}(\rm min)$\\
        \midrule
0 & 7.09 & 5.45 & 1.430 & 13.53 & 7.29 & 9.16 & 6.36 & 1.735 & 19.23 & 9.65\\
0.15 & 6.17 & 4.75 & 1.370 & 13.81 & 6.64 & 9.01 & 6.26 & 1.673 & 18.24 & 9.47\\
0.5 & 5.95 & 4.58 & 1.240 & 11.48 & 6.45 & 7.79 & 5.41 & 1.51 & 17.00 & 8.38\\
1.0 & 5.22 & 4.02 & 1.050 & 10.32 & 5.22 & 6.97 & 4.84 & 1.285 & 13.80 & 7.44\\
1.5 & 3.95 & 3.04 & 0.830 & 9.56 & 3.72 & 5.65 & 3.92 & 1.035 & 12.10 & 5.71\\
2.0 & 3.11 & 2.39 & 0.655 & 5.99 & 3.47 & 4.11 & 2.85 & 0.800 & 8.68 & 4.68\\

\toprule
    $Y_0=0.5$ & \multicolumn{5}{c}{$g_{14}=1.22 ~(0.5\,M_{\odot})$}& \multicolumn{5}{c}{$g_{14}=1.59~(1.0\,M_{\odot})$}\\
    \cmidrule(r){2-6}\cmidrule(r){7-11}
    $Q_{\rm b}(\rm MeV/u)$ & $f(\rm mHz)$ & $f_{\infty}(\rm mHz)$ & $\dot{m}/\dot{m}_{\rm Edd}$ & $L_{36}(\rm max)$ & $L_{36}(\rm min)$
      & $f(\rm mHz)$ & $f_{\infty}(\rm mHz)$ & $\dot{m}/\dot{m}_{\rm Edd}$  & $L_{36}(\rm max)$ & $L_{36}(\rm min)$\\
        \midrule
0 & 5.41 & 4.87 & 1.175 & 5.09 & 2.56 & 7.72 & 6.60 & 1.410 & 9.47 & 4.66\\
0.15 & 5.34 & 4.81 & 1.13 & 4.90 & 2.46 & 7.03 & 6.01 & 1.355 & 9.01 & 4.56\\
0.5 & 4.89 & 4.41 & 1.020 & 4.48 & 2.22 & 6.49 & 5.55 & 1.22 & 7.90 & 4.22\\
1.0 & 4.02 & 3.62 & 0.850 & 4.04 & 1.74 & 6.04 & 5.16 & 1.035 & 6.97 & 3.49\\
1.5 & 3.27 & 2.95 & 0.675 & 3.50 & 1.33 & 4.68 & 4.00 & 0.825 & 6.10 & 2.59\\
2.0 & 2.46 & 2.22 & 0.530 & 2.14 & 1.25 & 3.37 & 2.88 & 0.645 & 4.01 & 2.33\\
\toprule
    \multirow{2}{*}{} & \multicolumn{5}{c}{$g_{14}=2.16 ~(2.0\,M_{\odot})$}& \multicolumn{5}{c}{$g_{14}=2.86~(3.0\,M_{\odot})$}\\
    \cmidrule(r){2-6}\cmidrule(r){7-11}
     & $f(\rm mHz)$ & $f_{\infty}(\rm mHz)$ & $\dot{m}/\dot{m}_{\rm Edd}$ & $L_{36}(\rm max)$ & $L_{36}(\rm min)$
      & $f(\rm mHz)$ & $f_{\infty}(\rm mHz)$  & $\dot{m}/\dot{m}_{\rm Edd}$  & $L_{36}(\rm max)$ & $L_{36}(\rm min)$\\
        \midrule
0 & 10.96 & 8.43 & 1.740 & 17.20 & 8.48 & 13.89 & 9.65 & 2.110 & 25.95 & 10.71\\
0.15 & 9.75 & 7.50 & 1.675 & 16.22 & 8.35 & 13.55 & 9.41 & 2.025 & 21.98 & 11.59\\
0.5 & 9.16 & 7.38 & 1.51 & 14.28 & 7.76 & 12.53 & 8.70 & 1.840 & 20.44 & 10.26\\
1.0 & 7.65 & 5.88 & 1.285 & 12.51 & 6.49 & 10.96 & 7.61 & 1.558 & 17.76 & 8.58\\
1.5 & 6.29 & 4.84 & 1.030 & 11.09 & 4.86 & 8.82 & 6.13 & 1.28 & 12.99 & 7.85\\
2.0 & 4.80 & 3.69 & 0.810 & 7.36 & 4.33 & 6.59 & 4.58 & 0.990 & 10.24 & 5.93\\
        \bottomrule
    \end{tabular}
\end{table*}

\begin{table*}
    \centering
    \begin{tabular}{ccccccccccc}
    \toprule
    $Y_0=0.6$ & \multicolumn{5}{c}{$g_{14}=1.22 ~(0.5\,M_{\odot})$}& \multicolumn{5}{c}{$g_{14}=1.59~(1.0\,M_{\odot})$}\\
    \cmidrule(r){2-6}\cmidrule(r){7-11}
    $Q_{\rm b}(\rm MeV/u)$ & $f(\rm mHz)$ & $f_{\infty}(\rm mHz)$  & $\dot{m}/\dot{m}_{\rm Edd}$ & $L_{36}(\rm max)$ & $L_{36}(\rm min)$
      & $f(\rm mHz)$ & $f_{\infty}(\rm mHz)$ & $\dot{m}/\dot{m}_{\rm Edd}$  & $L_{36}(\rm max)$ & $L_{36}(\rm min)$\\
        \midrule
0 & 7.65 & 6.89 & 1.415 & 6.09 & 3.08 & 10.35 & 8.85 & 1.695 & 11.45 & 5.58\\
0.15 & 7.34 & 6.61 & 1.363 & 5.83 & 2.98 & 9.98 & 8.53 & 1.632 & 10.94 & 5.47\\
0.5 & 6.86 & 6.18 & 1.230 & 5.35 & 2.67 & 9.80 & 8.38 & 1.480 & 9.94 & 5.01\\
1.0 & 5.93 & 5.34 & 1.040 & 4.50 & 2.31 & 7.65 & 6.54 & 1.255 & 7.98 & 4.39\\
1.5 & 4.90 & 4.41 & 0.840 & 3.65 & 1.86 & 6.56 & 5.61 & 1.020 & 6.47 & 3.61\\
2.0 & 3.65 & 3.29 & 0.650 & 2.54 & 1.57 & 4.93 & 4.21 & 0.790 & 4.48 & 3.13\\

\toprule
    \multirow{2}{*}{} & \multicolumn{5}{c}{$g_{14}=2.16 ~(2.0\,M_{\odot})$}& \multicolumn{5}{c}{$g_{14}=2.86~(3.0\,M_{\odot})$}\\
    \cmidrule(r){2-6}\cmidrule(r){7-11}
     & $f(\rm mHz)$ & $f_{\infty}(\rm mHz)$ & $\dot{m}/\dot{m}_{\rm Edd}$ & $L_{36}(\rm max)$ & $L_{36}(\rm min)$
      & $f(\rm mHz)$ & $f_{\infty}(\rm mHz)$ & $\dot{m}/\dot{m}_{\rm Edd}$  & $L_{36}(\rm max)$ & $L_{36}(\rm min)$\\
        \midrule
0 & 15.15 & 11.65 & 2.090 & 22.67 & 9.28 & 19.61 & 13.62 & 2.540 & 28.92 & 13.72\\
0.15 & 13.77 & 10.59 & 2.015 & 19.52 & 10.04 & 19.16 & 13.31 & 2.450 & 26.52 & 13.92\\
0.5 & 13.89 & 10.68 & 1.830 & 17.76 & 9.23 & 19.61 & 13.62 & 2.220 & 24.69 & 12.41\\
1.0 & 11.74 & 9.03 & 1.543 & 15.72 & 7.55 & 15.72 & 10.92 & 1.891 & 19.03 & 11.59\\
1.5 & 9.36 & 7.2 & 1.265 & 11.57 & 6.77 & 12.92 & 8.97 & 1.545 & 15.75 & 9.45\\
2.0 & 7.61 & 5.85 & 0.985 & 8.95 & 5.27 & 9.86 & 6.85 & 1.201 & 12.22 & 7.44\\

\toprule
    $Y_0=0.7$ & \multicolumn{5}{c}{$g_{14}=1.22 ~(0.5\,M_{\odot})$}& \multicolumn{5}{c}{$g_{14}=1.59~(1.0\,M_{\odot})$}\\
    \cmidrule(r){2-6}\cmidrule(r){7-11}
    $Q_{\rm b}(\rm MeV/u)$ & $f(\rm mHz)$ & $f_{\infty}(\rm mHz)$ & $\dot{m}/\dot{m}_{\rm Edd}$ & $L_{36}(\rm max)$ & $L_{36}(\rm min)$
      & $f(\rm mHz)$ & $f_{\infty}(\rm mHz)$ & $\dot{m}/\dot{m}_{\rm Edd}$  & $L_{36}(\rm max)$ & $L_{36}(\rm min)$\\
        \midrule
0 & 11.11 & 10.01 & 1.688 & 7.66 & 3.47 & 15.02 & 12.84 & 2.026 & 13.42 & 6.86\\
0.15 & 10.48 & 9.44 & 1.626 & 6.92 & 3.60 & 14.75 & 12.61 & 1.950 & 12.71 & 6.60\\
0.5 & 10.16 & 9.15 & 1.475 & 6.33 & 3.26 & 13.33 & 11.39 & 1.76 & 11.16 & 6.17\\
1.0 & 8.46 & 7.62 & 1.240 & 5.61 & 2.65 & 11.49 & 9.82 & 1.5 & 9.74 & 5.20\\
1.5 & 7.03 & 6.33 & 1.010 & 4.48 & 2.20 & 9.69 & 8.28 & 1.225 & 7.90 & 4.28\\
2.0 & 5.18 & 4.67 & 0.780 & 3.15 & 1.83 & 7.12 & 6.09 & 0.940 & 6.18 & 3.29\\
\toprule
    \multirow{2}{*}{} & \multicolumn{5}{c}{$g_{14}=2.16 ~(2.0\,M_{\odot})$}& \multicolumn{5}{c}{$g_{14}=2.86~(3.0\,M_{\odot})$}\\
    \cmidrule(r){2-6}\cmidrule(r){7-11}
     & $f(\rm mHz)$ & $f_{\infty}(\rm mHz)$  & $\dot{m}/\dot{m}_{\rm Edd}$ & $L_{36}(\rm max)$ & $L_{36}(\rm min)$
      & $f(\rm mHz)$ & $f_{\infty}(\rm mHz)$  & $\dot{m}/\dot{m}_{\rm Edd}$  & $L_{36}(\rm max)$ & $L_{36}(\rm min)$\\
        \midrule
0 & 21.10 & 16.23 & 2.501 & 24.13 & 12.46 & 29.24 & 20.31 & 3.034 & 31.73 & 17.92\\
0.15 & 19.84 & 15.26 & 2.410 & 23.17 & 12.14 & 27.32 & 18.97 & 2.925 & 32.37 & 16.43\\
0.5 & 18.12 & 13.94 & 2.180 & 21.80 & 10.63 & 26.04 & 18.08 & 2.660 & 28.82 & 15.26\\
1.0 & 16.18 & 12.45 & 1.851 & 18.22 & 9.17 & 22.52 & 15.64 & 2.26 & 23.81 & 13.49\\
1.5 & 13.33 & 10.25 & 1.525 & 14.76 & 7.76 & 18.73 & 13.01 & 1.861 & 19.77 & 10.92\\
2.0 & 9.98 & 7.68 & 1.18 & 10.71 & 6.29 & 14.12 & 9.81 & 1.450 & 14.44 & 9.12\\

\toprule
    $Y_0=0.8$ & \multicolumn{5}{c}{$g_{14}=1.22 ~(0.5\,M_{\odot})$}& \multicolumn{5}{c}{$g_{14}=1.59~(1.0\,M_{\odot})$}\\
    \cmidrule(r){2-6}\cmidrule(r){7-11}
    $Q_{\rm b}(\rm MeV/u)$ & $f(\rm mHz)$ & $f_{\infty}(\rm mHz)$  & $\dot{m}/\dot{m}_{\rm Edd}$ & $L_{36}(\rm max)$ & $L_{36}(\rm min)$
      & $f(\rm mHz)$ & $f_{\infty}(\rm mHz)$ & $\dot{m}/\dot{m}_{\rm Edd}$  & $L_{36}(\rm max)$ & $L_{36}(\rm min)$\\
        \midrule
0 & 14.88 & 13.41 & 2.011 & 8.69 & 4.38 & 21.65 & 18.50 & 2.413 & 15.77 & 8.14\\
0.15 & 14.62 & 13.17 & 1.938 & 8.15 & 4.32 & 20.33 & 17.38 & 2.322 & 15.52 & 7.77\\
0.5 & 13.66 & 12.31 & 1.755 & 7.56 & 3.85 & 18.94 & 16.19 & 2.110 & 14.03 & 7.14\\
1.0 & 11.66 & 10.50 & 1.490 & 6.40 & 3.29 & 16.03 & 13.70 & 1.795 & 11.59 & 6.25\\
1.5 & 9.58 & 8.63 & 1.220 & 4.97 & 2.83 & 12.72 & 10.87 & 1.470 & 9.50 & 5.12\\
2.0 & 7.25 & 6.53 & 0.940 & 3.80 & 2.20 & 9.98 & 8.53 & 1.135 & 7.09 & 4.13\\
\toprule
    \multirow{2}{*}{} & \multicolumn{5}{c}{$g_{14}=2.16 ~(2.0\,M_{\odot})$}& \multicolumn{5}{c}{$g_{14}=2.86~(3.0\,M_{\odot})$}\\
    \cmidrule(r){2-6}\cmidrule(r){7-11}
     & $f(\rm mHz)$ & $f_{\infty}(\rm mHz)$ & $\dot{m}/\dot{m}_{\rm Edd}$ & $L_{36}(\rm max)$ & $L_{36}(\rm min)$
      & $f(\rm mHz)$ & $f_{\infty}(\rm mHz)$ & $\dot{m}/\dot{m}_{\rm Edd}$  & $L_{36}(\rm max)$ & $L_{36}(\rm min)$\\
        \midrule
0 & 31.45 & 24.19 & 2.980 & 26.79 & 15.88 & 40.65 & 28.23 & 3.624 & 40.27 & 20.13\\
0.15 & 26.88 & 20.68 & 2.875 & 27.70 & 14.48 & 39.68 & 27.56 & 3.495 & 38.18 & 19.69\\
0.5 & 24.15 & 18.58 & 2.593 & 30.67 & 10.78 & 35.46 & 24.63 & 3.160 & 38.88 & 16.02\\
1.0 & 21.93 & 16.87 & 2.215 & 21.02 & 11.30 & 29.76 & 20.67 & 2.710 & 29.32 & 15.62\\
1.5 & 18.12 & 13.94 & 1.82 & 18.07 & 8.94 & 26.46 & 18.38 & 2.220 & 24.66 & 12.63\\
2.0 & 14.01 & 10.78 & 1.420 & 12.87 & 7.57 & 19.38 & 13.46 & 1.74 & 16.23 & 11.56\\
        \bottomrule
    \end{tabular}
\end{table*}

\begin{table*}
    \centering
    \begin{tabular}{ccccccccccc}
    \toprule
    $Y_0=0.9$ & \multicolumn{5}{c}{$g_{14}=1.22 ~(0.5\,M_{\odot})$}& \multicolumn{5}{c}{$g_{14}=1.59~(1.0\,M_{\odot})$}\\
    \cmidrule(r){2-6}\cmidrule(r){7-11}
    $Q_{\rm b}(\rm MeV/u)$ & $f(\rm mHz)$ & $f_{\infty}(\rm mHz)$ & $\dot{m}/\dot{m}_{\rm Edd}$ & $L_{36}(\rm max)$ & $L_{36}(\rm min)$
      & $f(\rm mHz)$ & $f_{\infty}(\rm mHz)$ & $\dot{m}/\dot{m}_{\rm Edd}$  & $L_{36}(\rm max)$ & $L_{36}(\rm min)$\\
        \midrule
0 & 19.84 & 17.87 & 2.395 & 11.16 & 4.84 & 28.25 & 24.15 & 2.878 & 19.73 & 9.43\\
0.15 & 19.61 & 17.67 & 2.310 & 9.69 & 5.15 & 27.78 & 23.74 & 2.776 & 18.53 & 9.45 \\
0.5 & 18.52 & 16.68 & 2.090 & 8.42 & 4.89 & 26.04 & 22.26 & 2.52 & 16.54 & 8.59\\
1.0 & 16.03 & 14.44 & 1.788 & 7.04 & 4.29 & 22.22 & 18.99 & 2.15 & 14.27 & 7.41\\
1.5 & 13.44 & 12.11 & 1.466 & 6.45 & 3.28 & 17.73 & 15.15 & 1.760 & 11.99 & 5.76\\
2.0 & 10.22 & 9.21 & 1.140 & 4.60 & 2.70 & 13.77 & 11.77 & 1.370 & 9.03 & 4.85\\

\toprule
    \multirow{2}{*}{} & \multicolumn{5}{c}{$g_{14}=2.16 ~(2.0\,M_{\odot})$}& \multicolumn{5}{c}{$g_{14}=2.86~(3.0\,M_{\odot})$}\\
    \cmidrule(r){2-6}\cmidrule(r){7-11}
     & $f(\rm mHz)$ & $f_{\infty}(\rm mHz)$ & $\dot{m}/\dot{m}_{\rm Edd}$ & $L_{36}(\rm max)$ & $L_{36}(\rm min)$
      & $f(\rm mHz)$ & $f_{\infty}(\rm mHz)$ & $\dot{m}/\dot{m}_{\rm Edd}$  & $L_{36}(\rm max)$ & $L_{36}(\rm min)$\\
        \midrule
0 & 40.65 & 31.27 & 3.565 & 38.97 & 15.70 & 53.76 & 37.33 & 4.339 & 60.29 & 19.29\\
0.15 & 39.68 & 30.52 & 3.438 & 32.95 & 17.31 & 50.51 & 35.08 & 4.184 & 50.75 & 21.48\\
0.5 & 36.23 & 27.87 & 3.12 & 30.84 & 15.40 & 52.08 & 36.17 & 3.812 & 41.30 & 21.93\\
1.0 & 29.76 & 22.89 & 2.670 & 25.39 & 13.73 & 42.74 & 29.68 & 3.260 & 32.52 & 20.23\\
1.5 & 24.15 & 18.58 & 2.160 & 22.86 & 10.17 & 35.46 & 24.63 & 2.680 & 27.24 & 16.48\\
2.0 & 19.84 & 15.26 & 1.720 & 15.14 & 9.58 & 26.88 & 18.67 & 2.100 & 20.65 & 13.25\\

        \bottomrule
    \end{tabular}
    \label{tab:LX3630A}
\end{table*}

\begin{table*}
    \centering
    \caption{Physical quantities of oscillation frequency $f$, critical mass accretion rate $\dot{m}/\dot{m}_{\rm Edd}$ (the last value of the critical mass accretion rate in its narrow range), maxima luminosity $L_{36}(\rm max)$, minima luminosity $L_{36}(\rm min)$ with variation in metallicity $Z$. The gravity is set as $g_{14}=1.50$, which corresponds to a $1.5\,M_{\odot}$ NS from Togashi EoS.
    }
    \begin{tabular}{ccccccccc}   
    \toprule
      $Y_0=0.3$ & \multicolumn{4}{c}{$Q_{\rm b}=0\,(\rm MeV/u)$}& \multicolumn{4}{c}{$Q_{\rm b}=0.15 \,(\rm MeV/u)$}\\
    \cmidrule(r){2-5}\cmidrule(r){6-9}
       $Z$ & $f(\rm mHz)$ & $\dot{m}/\dot{m}_{\rm Edd}$ & $L_{36}(\rm max)$ & $L_{36}(\rm min)$
      & $f(\rm mHz)$ & $\dot{m}/\dot{m}_{\rm Edd}$  & $L_{36}(\rm max)$ & $L_{36}(\rm min)$\\
        \midrule
0.01 & 2.70 & 0.879 & 8.85 & 4.89 & 2.53 & 0.850 & 8.76 & 4.66\\
0.03 & 2.64 & 0.850 & 8.72 & 4.66 & 2.71 & 0.812 & 8.82 & 4.29\\
0.05 & 2.59 & 0.812 & 8.52 & 4.37 & 2.56 & 0.777 & 7.92 & 4.32\\
0.07& 2.51 & 0.773 & 8.25 & 4.13 & 2.29 & 0.731 & 7.79 & 3.89\\
0.1& 2.32 & 0.703 & 7.33 & 3.87 & 2.17 & 0.653 & 6.95 & 3.50\\
 \toprule
     & \multicolumn{4}{c}{$Q_{\rm b}=0.5\,(\rm MeV/u)$}& \multicolumn{4}{c}{$Q_{\rm b}=1.0 \,(\rm MeV/u)$}\\
    \cmidrule(r){2-5}\cmidrule(r){6-9}
     & $f(\rm mHz)$ & $\dot{m}/\dot{m}_{\rm Edd}$ & $L_{36}(\rm max)$ & $L_{36}(\rm min)$
      & $f(\rm mHz)$ & $\dot{m}/\dot{m}_{\rm Edd}$  & $L_{36}(\rm max)$ & $L_{36}(\rm min)$\\
        \midrule
0.01 & 2.41 & 0.760 & 7.98 & 4.14 & 1.95 & 0.642 & 6.68 & 3.50\\
0.03 & 2.30 & 0.726 & 7.55 & 3.97 & 1.89 & 0.595 & 5.99 & 3.40\\
0.05 & 2.17 & 0.680 & 7.25 & 3.65 & 1.77 & 0.535 & 5.38 & 3.02\\
0.07& 2.07 & 0.625 & 6.73 & 3.33 & 1.34 & 0.450 & 4.38 & 2.63\\
0.1 & 1.70 & 0.515 & 5.32 & 2.85 & - & - & - & -\\

 \toprule
     & \multicolumn{4}{c}{$Q_{\rm b}=1.5\,(\rm MeV/u)$}& \multicolumn{4}{c}{$Q_{\rm b}=2.0 \,(\rm MeV/u)$}\\
    \cmidrule(r){2-5}\cmidrule(r){6-9}
  & $f(\rm mHz)$ & $\dot{m}/\dot{m}_{\rm Edd}$ & $L_{36}(\rm max)$ & $L_{36}(\rm min)$
      & $f(\rm mHz)$ & $\dot{m}/\dot{m}_{\rm Edd}$  & $L_{36}(\rm max)$ & $L_{36}(\rm min)$\\
        \midrule
0.01 & 1.56 & 0.516 & 5.20 & 2.91 & 1.19 & 0.385 & 3.85 & 2.22\\
0.03 & 1.43 & 0.455 & 4.10 & 2.84 & 0.77 & 0.265 & 2.52 &1.61\\
0.05 & 1.03 & 0.335 & 3.25 & 1.98 & - & - & - & \\
0.07& - & - & - & - & - & - & - & -\\
0.1 & - & - & - & - & - & - & - & -\\

        \bottomrule
    \end{tabular}
    \label{tab:LX3630}
\end{table*}


\bsp	
\label{lastpage}
\end{document}